\documentclass{aa}
\usepackage{epsfig}

\def\etal{{\rm et al.\thinspace}}
\def\eg{{\rm e.g.}}

\def\ie{{\rm i.e.\ }}
\def\cf{{\rm cf.\ }}

\def\spose#1{\hbox to 0pt{#1\hss}}
\def\ltsimm{\mathrel{\spose{\lower 3pt\hbox{$\sim$}}
	\raise 2.0pt\hbox{$<$}}}
\def\gtsimm{\mathrel{\spose{\lower 3pt\hbox{$\sim$}}
	\raise 2.0pt\hbox{$>$}}}

\def\km{{\rm\thinspace km}}
\def\cm{{\rm\thinspace cm}}

\def\s{{\rm\thinspace s}}

\def\g{{\rm\thinspace g}}

\def\kmps{\hbox{${\rm\km\s^{-1}\,}$}}
\def\erg{{\rm\thinspace erg}}

\def\ergpcm2ps{\hbox{${\rm\erg\cm^{-2}\s^{-1}\,}$}}
\def\pcm3{\hbox{${\rm\cm^{-3}\,}$}}
\def\gpcm3{\hbox{${\rm\g\cm^{-3}\,}$}}

\begin{document}
   
\title{The Shocking Properties of Supersonic Flows: Dependence of the Thermal
Overstability on $M$, $\alpha$, and $T_{\rm c}/T_{0}$}

\author{J.M. Pittard\inst{1}, M.S. Dobson\inst{1}, R.H. Durisen\inst{2}, 
J.E. Dyson\inst{1}, T.W. Hartquist\inst{1}, J.T. O'Brien\inst{1}}

\institute{School of Physics and Astronomy, The University of Leeds, 
        Woodhouse Lane, Leeds, LS2 9JT, UK
\and
Department of Astronomy, Indiana University, Swain Hall West 319, 727 East 3rd St., Bloomington, 47405, USA
\\}

\offprints{J.M. Pittard, \email{jmp@ast.leeds.ac.uk}}

\date{Accepted 8$^{th}$ April, 2005}

\abstract{We present hydrodynamical calculations of radiative shocks
with low Mach numbers and find that the well-known global
overstability can occur if the temperature exponent ($\alpha$) of the
cooling is sufficiently negative. We find that the stability of
radiative shocks increases with decreasing Mach number, with the
result that $M=2$ shocks require $\alpha \ltsimm -1.2$ in order to be
overstable. Such values occur within a limited temperature range of
many cooling curves.  We observe that Mach numbers of order 100 are
needed before the strong shock limit of $\alpha_{\rm cr} \approx 0.4$
is reached, and we discover that the frequency of oscillation of the
fundamental mode also has a strong Mach number dependence. We find
that feedback between the cooling region and the cold dense layer (CDL)
further downstream is a function of Mach number, with stronger
feedback and oscillation of the boundary between the CDL
and the cooling region occuring at lower Mach numbers. This feedback
can be quantified in terms of the reflection coefficient of sound
waves, and in those cases where the cooling layer completely
disappears at the end of each oscillation cycle, the initial velocity
of the waves driven into the upstream pre-shock flow and into the
downstream CDL, and the velocity of the the boundary between the CDL and
the cooling layer, can be understood in terms of the solution to the
Riemann problem. An interesting finding is that the stability properties 
of low Mach number shocks can be dramatically altered if the shocked gas is
able to cool to temperatures less than the pre-shock value (\ie when
$\chi < 1$, where $\chi$ is the ratio of the temperature of the cold
dense layer to the pre-shock temperature).  In such circumstances, low
Mach number shocks have values of $\alpha_{\rm cr}$ which are
comparable to values obtained for higher Mach number shocks when $\chi
= 1$. For instance, $\alpha_{\rm cr}=-0.1$ when $M=2$ and $\chi=0.1$,
comparable to that when $M=10$ and $\chi=1$.  Thus, it is probable
that low Mach number astrophysical shocks will be overstable in a
variety of situations.  We also explore the effect of different
assumptions for the initial hydrodynamic set up and the type of
boundary condition imposed downstream, and find that the properties of
low Mach number shocks are relatively insensitive to these issues.
The results of this work are relevant to astrophysical shocks with low
Mach numbers, such as supernova remnants (SNRs) immersed in a hot
interstellar medium (\eg, within a starburst region), and shocks in
molecular clouds, where time-dependent chemistry can lead to
overstability.
}

\authorrunning{Pittard \etal}
\titlerunning{The Shocking Properties of Supersonic Flows}

\maketitle

\label{firstpage}

\section{Introduction}
\label{sec:intro}
Radiative shocks were first shown to be susceptible to a cooling
instability by Falle (\cite{F1975}) and McCray \etal
(\cite{MKS1975}), and to oscillations from a global overstability by
Langer \etal (\cite{L1981}), and have since been extensively examined
(see Pittard \etal \cite{P2003} and references therein). A central
finding is that the stability of a radiative shock deteriorates as the
Mach number increases (Strickland \& Blondin \cite{SB1995}), since the
higher density of the cold gas layer behind radiative shocks of higher
Mach number acts increasingly like a reflecting wall, and reflects
incident waves with virtually the same amplitude. In contrast,
incident waves are reflected with much decreased amplitude in lower
Mach number shocks since the lower density cold gas layer acts more
like a cushion. 

In most investigations of the radiative overstability, the temperature
dependent cooling function is approximated as $\Lambda(T) =
\Lambda_{0} T^{\alpha}$, with the rate of energy loss per unit
volume, $\dot{E} = n^{2}\Lambda(T)$. The radiative overstability is
known to depend upon the value of $\alpha$: if $\alpha$ is greater
than some critical value, $\alpha_{\rm cr}$, the system is stable to
the fundamental mode of oscillation (though it may still be unstable
to higher overtones). Previous numerical work has shown that for high
Mach numbers shocks $\alpha_{\rm cr} \approx 0.4$ (\eg, Imamura \etal
\cite{IWD1984}; Strickland \& Blondin \cite{SB1995}), in good
agreement with the linear stability analysis of Chevalier \& Imamura
(\cite{CI1982}). The first overtone mode is stabilized when $\alpha
\gtsimm 0.8$. Further work has revealed that non-radial transverse
modes may be unstable at values of $\alpha$ which are stable to radial
modes (\cf Bertschinger \cite{B1986}), and that non-radial effects are
greater at higher densities, perhaps because of additional types of
instabilities (see Blondin \etal \cite{BWBR1998}).

To date, most investigations of the global overstability of radiative
shocks have been concerned with high Mach number shocks ($M \geq
5$). While simulations of decelerating radiative shocks in SNRs (\eg,
Kimoto \& Chernoff \cite{KC1997}; Blondin \etal \cite{BWBR1998}) and
wind-blown bubbles (Walder \& Folini \cite{WF1996}) ``scan'' a wide
range of Mach numbers, the ISM densities and temperatures
considered in these works resulted in SNRs with slow (due to ISM
density) low Mach number (due to ISM temperature) forward shocks, with
post-shock temperatures (slightly higher than the ISM temperature)
which lie in a range on the cooling curve where $\alpha$ is much
larger than zero.  Thus there is no instability. However, instability
may occur if the temperature of the ISM is larger, such that the
cooling function then has a significantly steep temperature
dependence, though there is no information in the current literature
on the value of $\alpha_{\rm cr}$ for low Mach number shocks.  The
ISM temperature\footnote{Usual practice is to set the
floor-temperature of the simulation and $T_{\rm ISM}$ to the same
value. Most of the simulations in this paper adopt this assumption,
though the effect of letting the gas cool below the pre-shock
temperature is examined in Sec.~\ref{sec:tcdl_ne_tamb}, where we find
that there are dramatic changes in the resulting properties when the
Mach number is low.} assumed in existing papers is typically $\sim
10^{4}\;$K.

In this paper we therefore examine the overstability of low Mach
shocks, extending previous numerical work to Mach numbers below
5. Since the boundary conditions imposed in numerical models of
radiative shocks can also affect their stability (Strickland \&
Blondin \cite{SB1995}; Saxton \cite{S2002}), we investigate the nature
of the instability as the boundary conditions and initial conditions
are varied. In Sec.~\ref{sec:properties} we summarize the properties
of low Mach number radiative shocks. Hydrodynamical models of the
overstability are presented in Sec.~\ref{sec:hydro} where we also
examine how feedback between the cold, dense layer (CDL) downstream of
the cooling layer, and the cooling region depends on the Mach
number. Additional results obtained when we allow the shocked gas to
cool to a temperature lower than that of the pre-shock flow are
noted. We discuss applications of our work in Sec.~\ref{sec:discuss},
and Sec.~\ref{sec:summary} summarizes the paper.

\section{Properties of low Mach number shocks}
\label{sec:properties}
\begin{figure}[ht]
\begin{center}
\psfig{figure=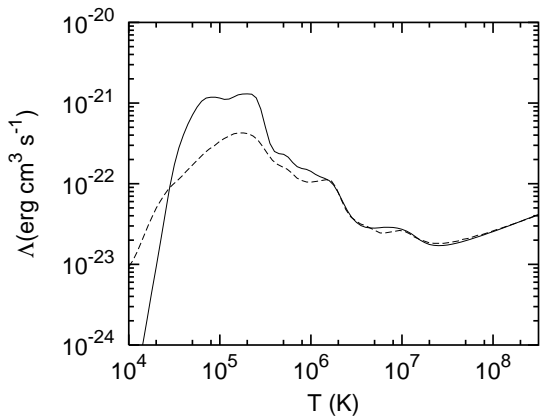,width=8.0cm}
\end{center}
\caption[]{Cooling curves for material of solar abundance assuming
collisional ionization equilibrium (solid) or non-equilibrium
ionization (dashed). The former was calculated using the MEKAL thermal
plasma code (Mewe \etal \cite{M1985}; Kaastra \cite{K1992}) 
distributed in XSPEC (v11.2.0), while the latter was taken from data in
Sutherland \& Dopita (\cite{SD1993}). These curves are normalized so that 
the net cooling rate per unit volume, 
$\dot{E} = n_{\rm e} n_{\rm i} \Lambda(T)$, where $n_{\rm e}$ is
the electron number density and $n_{\rm i}$ is the total number density
of all of the ions.}
\label{fig:cool_alpha1}
\end{figure}

\begin{figure}[ht]
\begin{center}
\psfig{figure=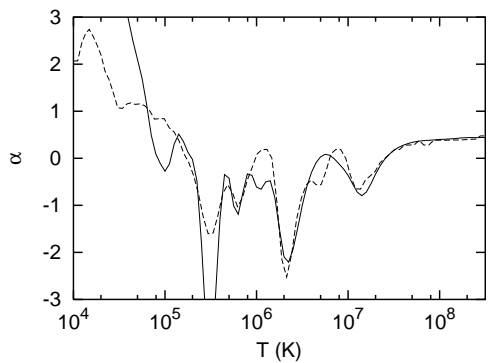,width=7.0cm}
\end{center}
\caption[]{The slope, $\alpha$, of the respective cooling curves
in Fig.~\ref{fig:cool_alpha1} (CIE - solid; NEI - dashed).}
\label{fig:cool_alpha2}
\end{figure}

The approximation that $\Lambda(T) = \Lambda_{0} T^{\alpha}$ is
particularly appropriate for the low Mach number flows which are the
subject of this work because the shocked gas cools through a
relatively narrow range of temperatures. Some properties of radiative
shocks are tabulated in Table~\ref{tab:properties}. The ratio of post-
to pre-shock density is given by
\begin{equation}
\label{eq:rhojump}
\frac{\rho_{\rm s}}{\rho_{\rm 0}} = \frac{(\gamma+1)M^{2}}{(\gamma-1)M^{2}+2},
\end{equation}
the ratio of post- to pre-shock temperature is given by
\begin{equation}
\label{eq:tempjump}
\frac{T_{\rm s}}{T_{\rm 0}} = \frac{[2\gamma M^{2}-(\gamma-1)]
[(\gamma-1) M^{2}+2]}{[(\gamma+1)M]^{2}},
\end{equation}
and the ratio of final to pre-shock density is
\begin{equation}
\label{eq:rhocompress}
\frac{\rho_{\rm c}}{\rho_{\rm 0}} = \gamma M^{2}
\end{equation} 
if the gas cools to its initial temperature, $T_{\rm 0}$.  In
Eqs.~\ref{eq:rhojump}-\ref{eq:rhocompress} the subscript ``${\rm
s}$'' denotes quantities immediately behind the shock front, while the
subscript ``${\rm 0}$'' indicates pre-shock quantities. The subscript
``${\rm c}$'' indicates quantities associated with the CDL. One
consequence of the reduced ratio of $T_{\rm s}/T_{\rm 0}$ in low Mach
number shocks is that $\alpha$ can obtain fairly extreme negative
values when the local slope of the cooling curve becomes very steep.
This is demonstrated in Figs.~\ref{fig:cool_alpha1} and~\ref{fig:cool_alpha2} 
where we show two
example cooling curves and the slope $\alpha$.  In both curves there
are two distinct temperature ranges where $\alpha \ltsimm -2$, and in
the cooling curve where collisional ionization equilibrium is enforced
we find that $\alpha \approx -4$ when $T \approx 3 \times
10^{5}\;$K. We further note that time-dependent chemistry behind
shocks in molecular clouds can also result in effective values of
$\alpha$ which are extremely negative (Smith \& Rosen \cite{SR2003}).

\begin{table}
\begin{center}
\caption{Properties of a radiative shock for gas with a ratio
of specific heats, $\gamma=5/3$.}
\label{tab:properties}
\begin{tabular}{llll}
\hline
$M$ & $\rho_{\rm s}/\rho_{\rm 0}$ & $\rho_{\rm c}/\rho_{\rm 0}$ 
& $T_{\rm s}/T_{\rm 0}$ \\
\hline
1.4 & 1.58 & 3.27 & 1.39 \\
2.0 & 2.29 & 6.67 & 2.08 \\
3.0 & 3.00 & 15.0 & 3.66 \\
5.0 & 3.57 & 41.7 & 8.68 \\
10.0 & 3.88 & 167 & 32.1 \\
20.0 & 3.97 & 667 & 126 \\
40.0 & 3.99 & 2667 & 501 \\
\hline
\end{tabular}
\end{center}
\end{table}

The cooling length of a radiative shock can be parameterized as
\begin{equation}
\label{eq:Lc}
L_{\rm c} = x_{\rm b} L_{\rm c}' \equiv x_{\rm b} \frac{v_{\rm s} 
\bar{m} k T_{\rm s}}{\rho_{\rm s} \Lambda_{0} T_{\rm s}^{\alpha}},
\end{equation}
where $\rho$, $v$ and $T$ are the fluid mass density, velocity, and
temperature, respectively, $\bar{m}$ is the mean mass
per particle ($\rho = \bar{m} n$ where $n$ is the number density) 
and $k$ is Boltzmann's constant.
$L_{\rm c}$ is defined as the distance behind the shock that
a parcel of gas travels before cooling to the pre-shock temperature, 
at which point cooling is usually assumed to turn off.
$x_{\rm b}$ is a factor of order unity. The structure of the shock
and the value of $x_{\rm b}$ can be determined by the method noted
in Strickland \& Blondin (\cite{SB1995}). Values for $L_{\rm c}$ and
$x_{\rm b}$ are given in Table~\ref{tab:coollength} for the case 
where $\bar{m}$, $k$ and $\Lambda_{0}$ are equal to unity. 
This assumption is also adopted for the numerical work which follows.

\begin{table}
\begin{center}
\caption{Cooling length of a radiative shock for gas with a ratio
of specific heats, $\gamma=5/3$, and cooling exponent, $\alpha=0$.  
$L_{\rm c}'$ in appropriate units can be obtained by multiplying by
the numerical values of $\bar{m} k/\Lambda_{0}$ (note that in this
work we also specify that $\rho_{0}=1$ and $v_{0}=1$).
There is convergence in the value of $x_{\rm b}$ as $M$ increases.}
\label{tab:coollength}
\begin{tabular}{llll}
\hline
$M$ & $L_{\rm c}'$ &	$x_{\rm b}$ & $L_{\rm c}$ \\   
\hline
 1.40 & 0.171     & 0.545 &    9.294E-02 \\
 2.00 & 5.966E-02 & 0.685 &    4.087E-02 \\
 3.00 & 2.716E-02 & 0.748 &    2.032E-02 \\
 5.00 & 1.633E-02 & 0.769 &    1.256E-02 \\
 10.0 & 1.278E-02 & 0.776 &    9.917E-03 \\
 20.0 & 1.198E-02 & 0.777 &    9.307E-03 \\
 40.0 & 1.178E-02 & 0.778 &    9.167E-03 \\
\hline
\end{tabular}
\end{center}
\end{table}

\section{Hydrodynamical Simulations}
\label{sec:hydro}
In this section we describe the results obtained with the hydrodynamical 
code VH-1 (see Blondin \etal \cite{BKFT1990}). 
VH-1 uses a Lagrangian formulation of the piecewise 
parabolic method (PPM; see Colella \& Woodward \cite{CW1984}), 
where the spatial representation of the fluid variables is third order 
accurate. After each timestep the conserved quantities are remapped 
onto the original grid. Recent work in which this scheme was used 
for investigations of the overstability of radiative shocks includes
Strickland \& Blondin (\cite{SB1995}), Sutherland \etal (\cite{SBD2003}), 
and Pittard \etal (\cite{P2003}, \cite{P2004}).

Radiative cooling is implemented via operator splitting and an 
unconditionally stable implicit scheme (see Strickland \& Blondin 
\cite{SB1995}). At unresolved phases between hot gas and denser cold
gas, the rate of energy loss of a given hydrodynamical cell due to 
radiative cooling is restricted to the minimum from the neighbouring
cells. This procedure allows the rapid cooling layer at the rear
of radiative shocks to be ``resolved'' with relatively few hydrodynamical
cells, and provides the advantage of a significant speed-up to computational
times. Resolution tests without this restriction show that the 
oscillation frequency and amplitude of the overstable shock converge 
towards the results obtained with this restriction as the numerical
resolution is increased (see also Sutherland \etal \cite{SBD2003}).  
We have performed additional tests to check that the behaviour of 
the system converges with increasing numerical resolution when 
$\alpha = \alpha_{\rm cr}$.

While a substantial body of work on the overstability of high 
Mach number shocks exists, it is difficult to make comparisons between results
since the adopted boundary conditions often differ. It has been shown that
the boundary conditions can play an important role in determining the
stability of the shock (see Strickland \& Blondin \cite{SB1995}). 
Therefore, we have examined the effects of 3 different types of initial and
boundary conditions. Results from each are presented in turn, where 
we prevent shocked gas from cooling below the pre-shock temperature.
Simulations where we allow the gas to cool further are presented in
Sec.~\ref{sec:tcdl_ne_tamb}.

\subsection{Impulsive shock generation}
\label{sec:iso_imp}
\begin{figure}[ht]
\begin{center}
\psfig{figure=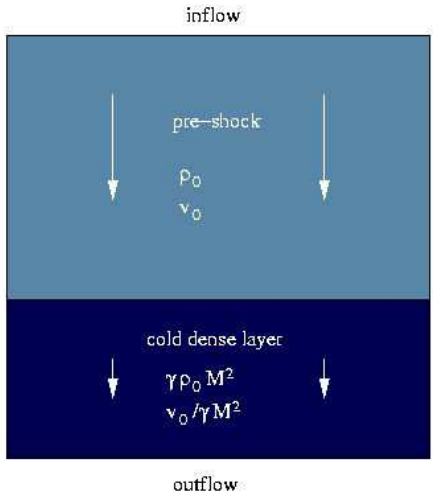,width=6.0cm}
\end{center}
\caption[]{Schematic of the grid set up for implusive shock generation.
Supersonic flow with density $\rho_{\rm 0}$ and flow speed $v_{\rm 0}$ 
collides with a subsonic CDL with density $\gamma M^{2} \rho_{\rm 0}$ and
flow speed $v_{\rm 0}/\gamma M^{2}$. The upstream boundary condition is inflow,
while the downstream boundary condition is outflow. All simulations are
1-D.}
\label{fig:isoimp}
\end{figure}

In this set of simulations we run a supersonic flow into a pre-existing cold, 
dense layer of gas, as shown schematically in Fig.~\ref{fig:isoimp}. The 
density and velocity of the supersonic flow are set to unity, and its pressure 
and temperature are set to $1/\gamma M^{2}$. The density of the cold 
layer is set to $\gamma M^{2}$, and its thermal pressure is set to 
unity (\ie equal to the ram pressure of the oncoming flow). The velocity of
gas within the cold layer is set to $1/\gamma M^{2}$. The total pressure 
(thermal plus ram) within the supersonic flow is equal to that within the 
cold dense slab. An inflow boundary exists in the supersonic flow, while
an outflow boundary is specified at the edge of the CDL.
All simulations are one-dimensional and use $\gamma=5/3$.

We set up the grid such that the width of each hydrodynamical cell is
constant in the supersonic flow and for several cooling lengths
($L_{\rm c}$) into the CDL, after which the width of each
cell increases by 3\% relative to its neighbour. This enables the
outflow boundary to be positioned so far downstream that it does not
affect the evolution of the overstability (\ie the shock driven into
the CDL does not reach the outflow boundary by the time
we stop our simulations). Our simulations typically contain $\sim
300-500$ cells within the cooling length of the shock.

\begin{figure}[ht]
\begin{center}
\psfig{figure=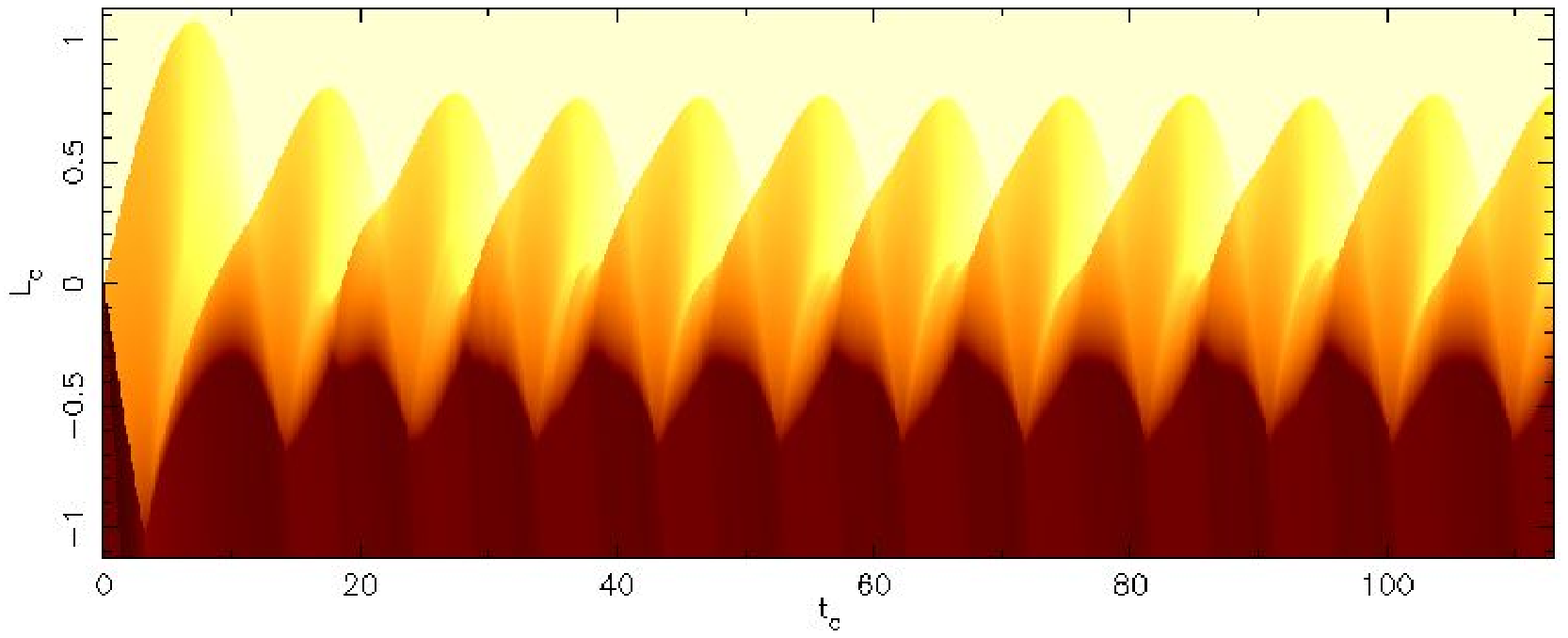,width=9.0cm,angle=0}
\psfig{figure=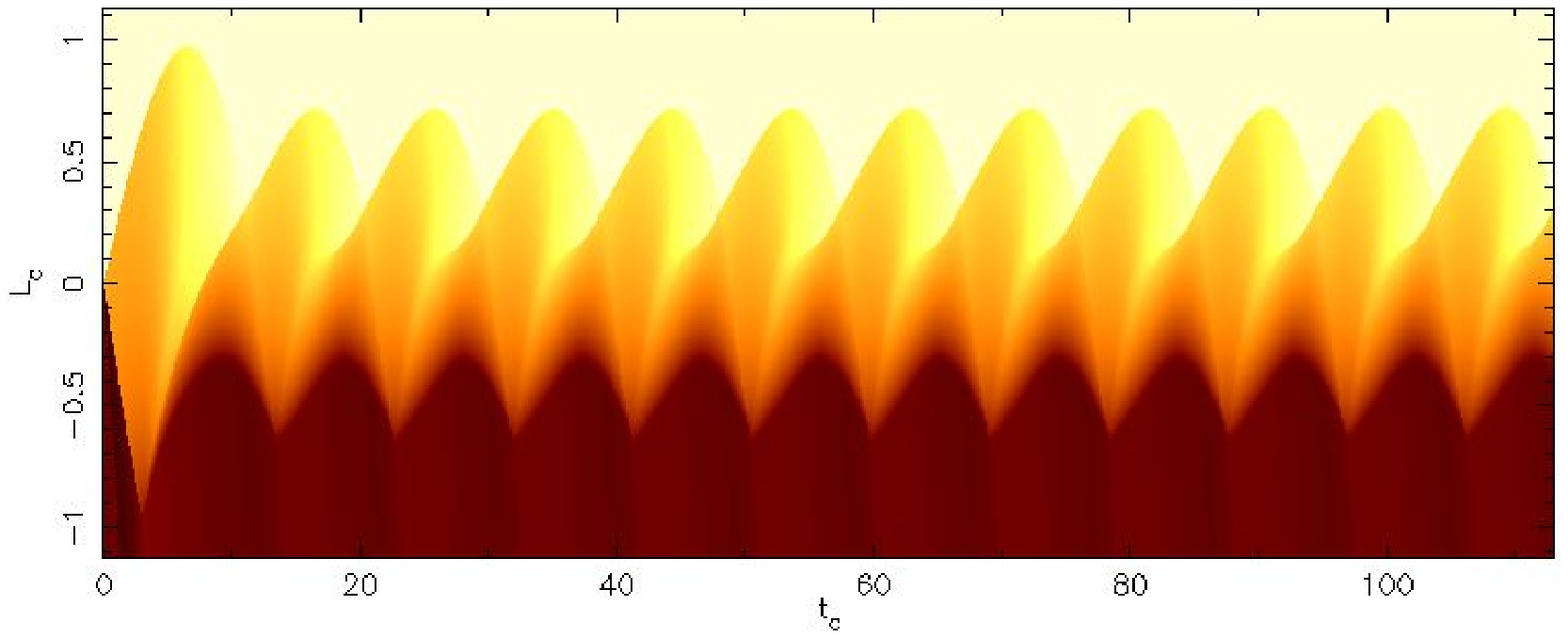,width=9.0cm,angle=0}
\psfig{figure=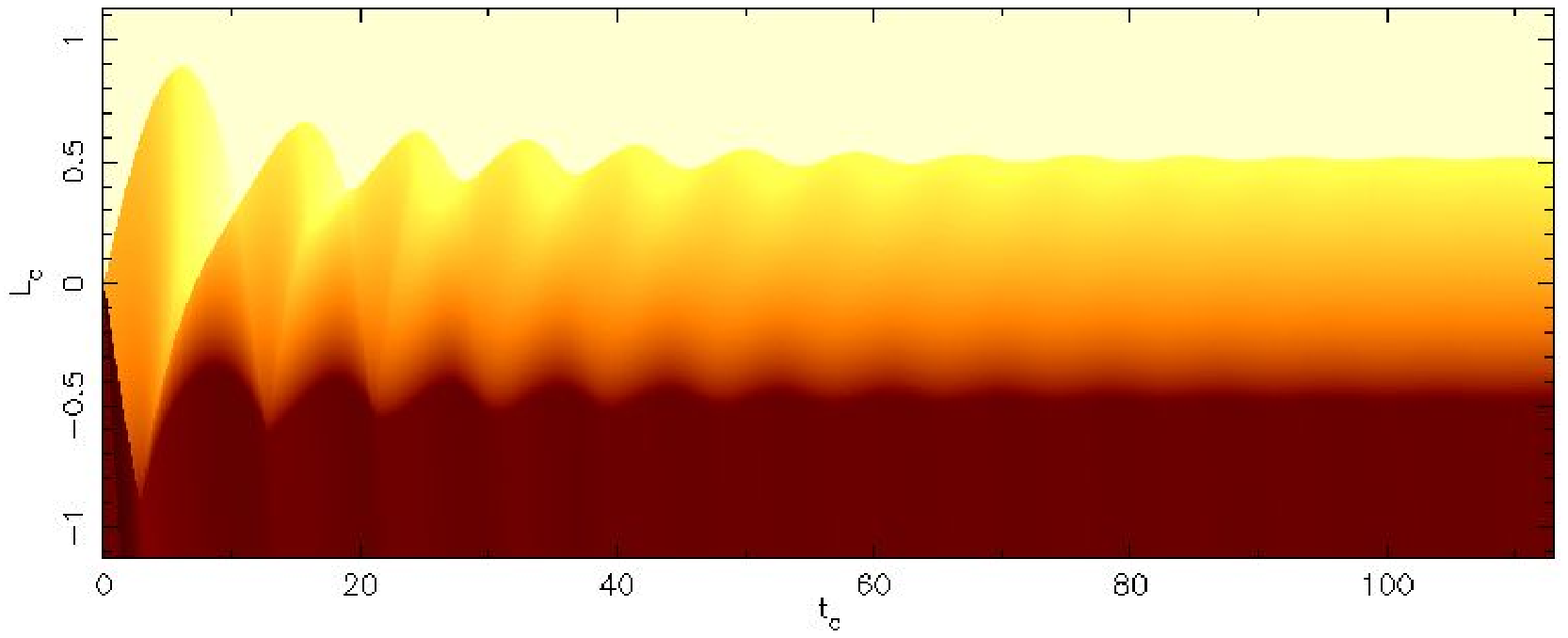,width=9.0cm,angle=0}
\end{center}
\caption[]{Time-space diagrams of the density evolution of a
1-dimensional Mach 1.4 radiative shock with different cooling
exponents $\alpha$.  Supersonic flow enters the grid from the top, and
the cooled postshock gas flows off the grid at the bottom. Lighter
shades indicate lower densities.  Distances are marked in units of
$L_{\rm c}$ (the value of this is different in each panel - see
Eq.~\ref{eq:Lc} and Table~\ref{tab:coollength}) while time is shown in
units of $L_{\rm c}/v_{\rm 0}$ ($v_{\rm 0} = 1$). The value of
$\alpha$ is $-2.5$, $-2.0$, and $-1.5$ in the top, middle, and bottom panels
respectively.}
\label{fig:imp_mach1.4}
\end{figure}

\begin{figure}[ht]
\begin{center}
\psfig{figure=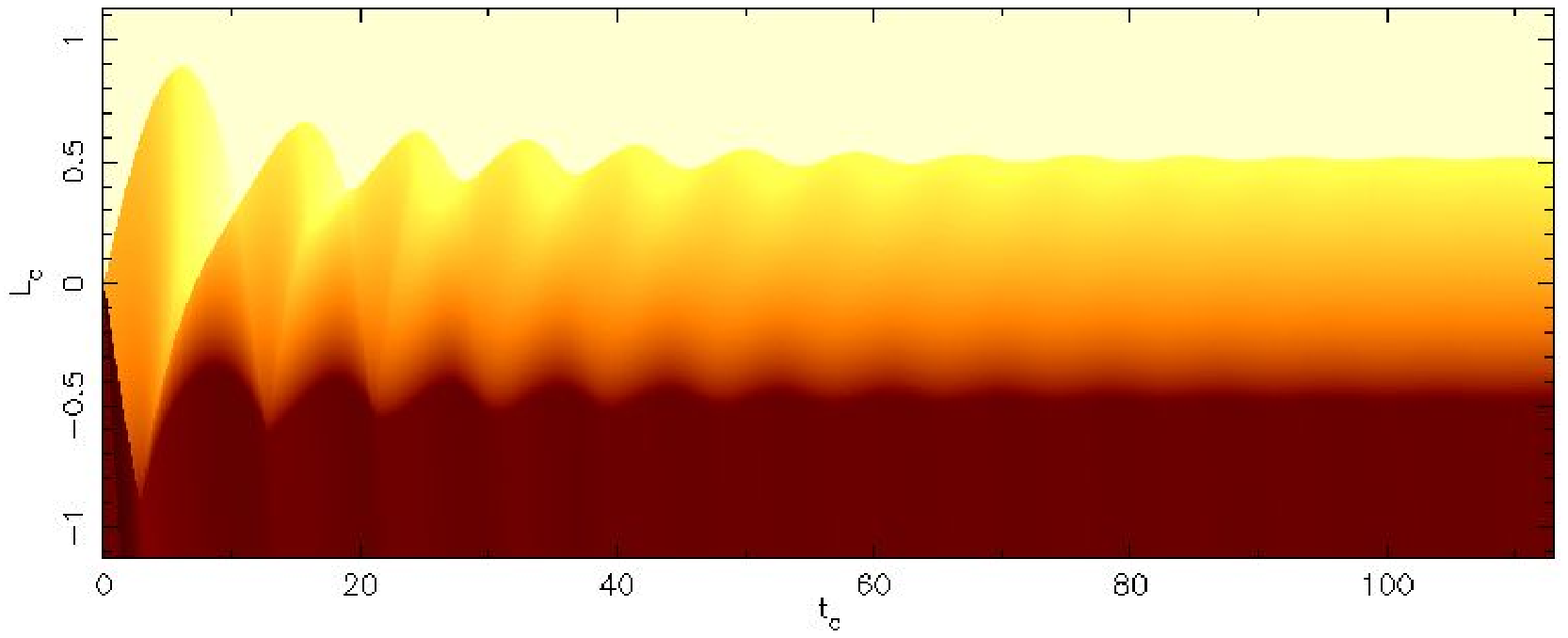,width=9.0cm,angle=0}
\psfig{figure=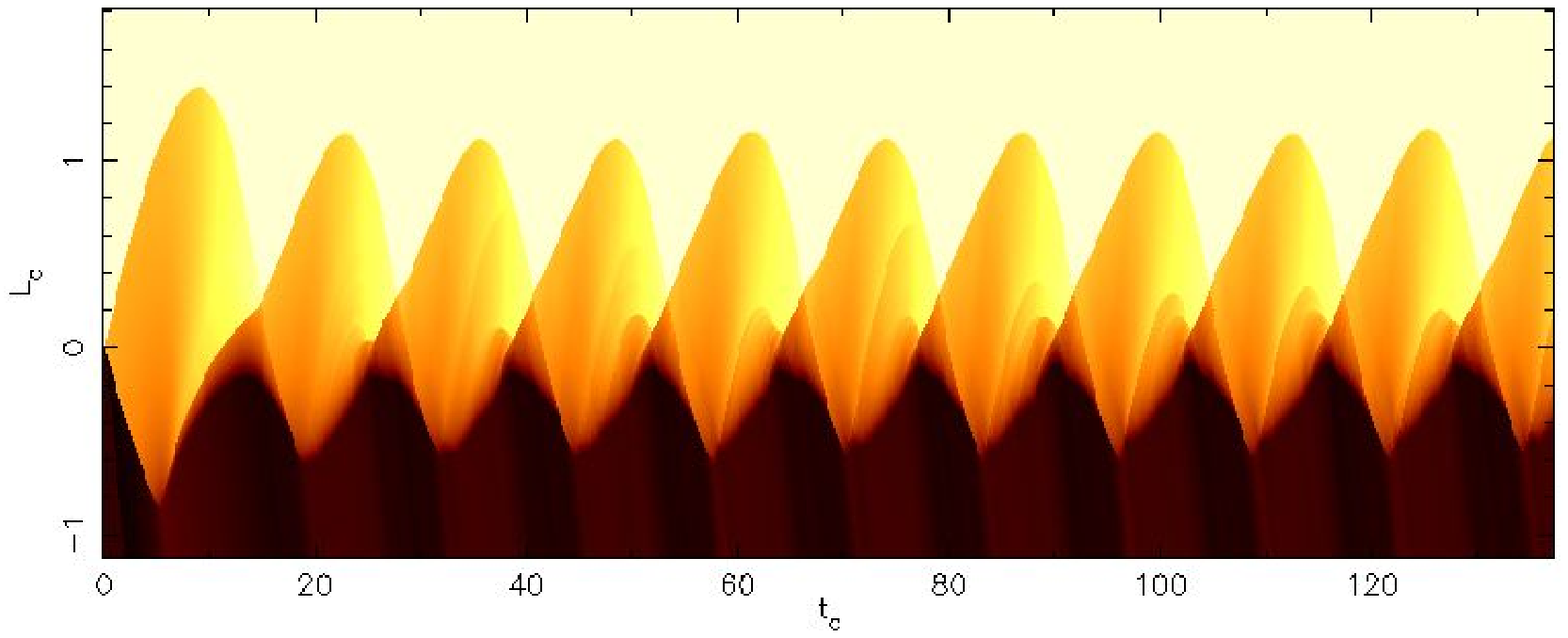,width=9.0cm,angle=0}
\psfig{figure=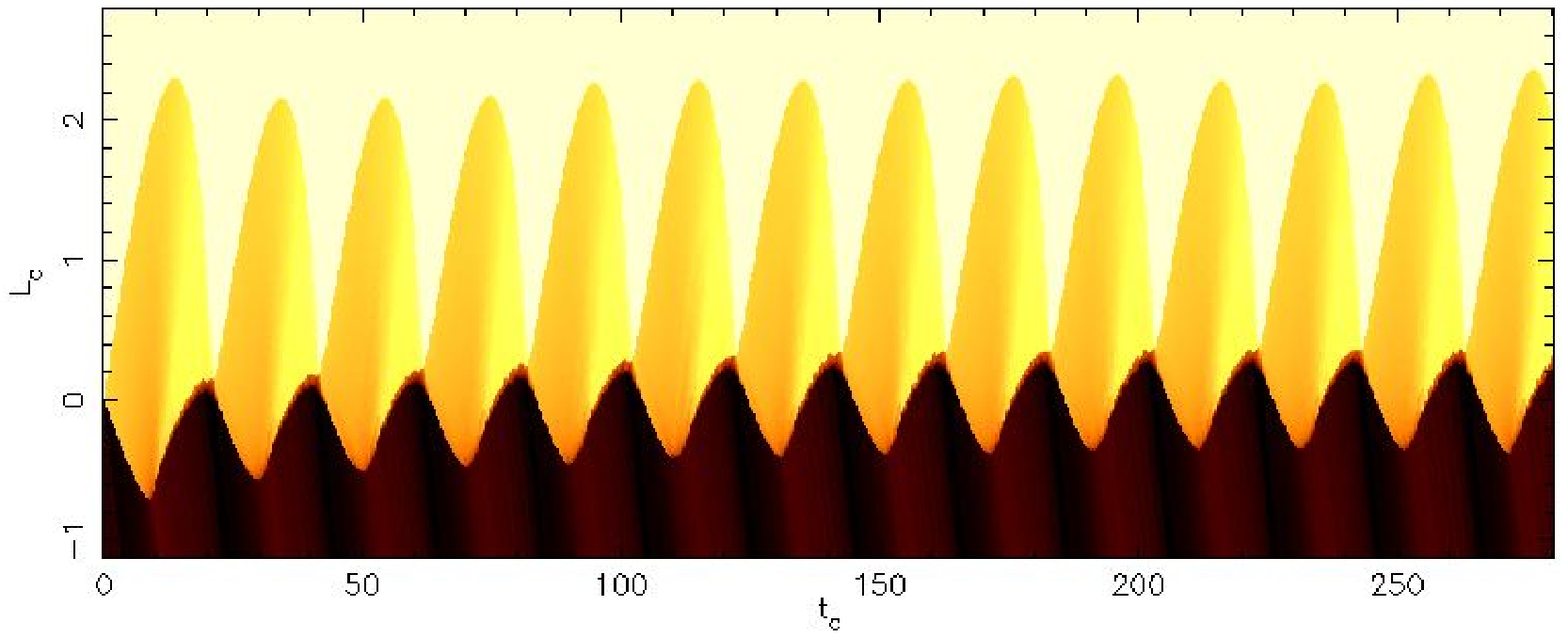,width=9.0cm,angle=0}
\psfig{figure=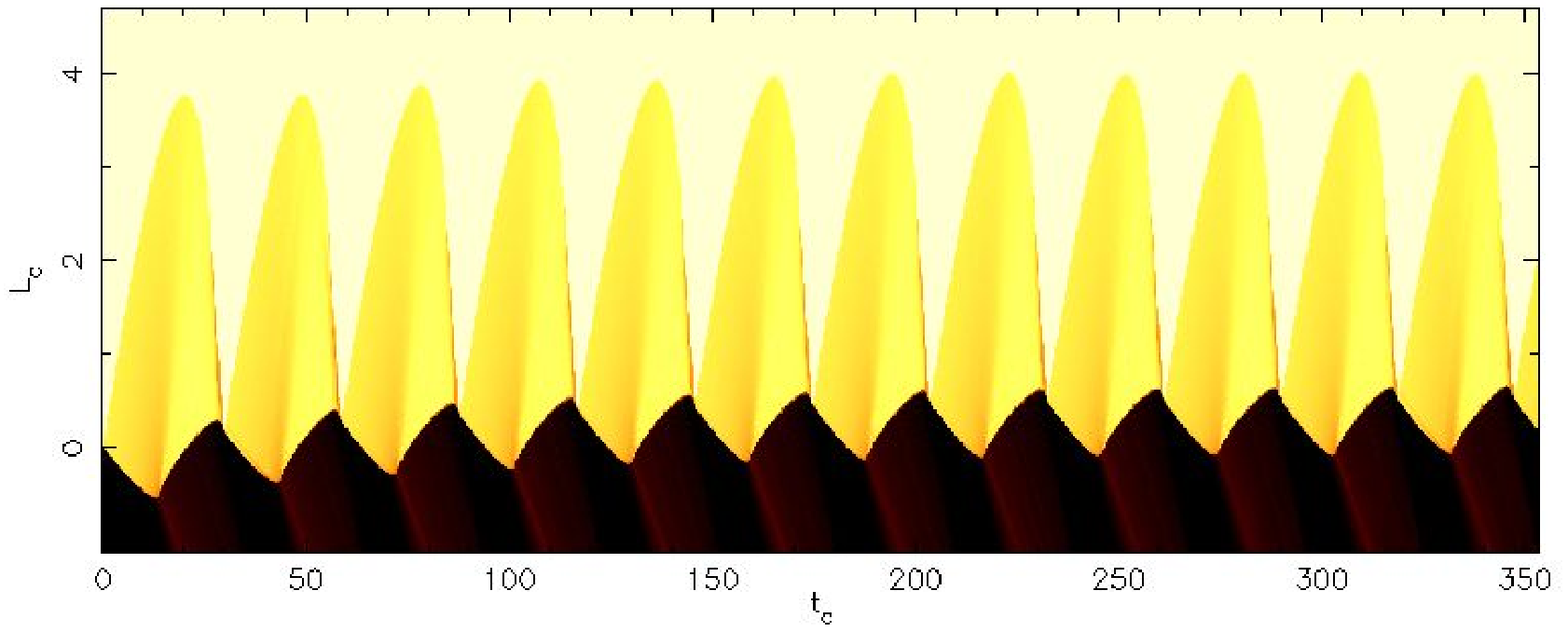,width=9.0cm,angle=0}
\end{center}
\caption[]{Time-space diagrams of the density evolution of a 1-dimensional
radiative shock with cooling exponents $\alpha=-1.5$ and different Mach 
numbers. $M$ is varied from 1.4, 2, 3 and 5 from the top to bottom panels.
Note that the axis and density scaling changes from panel to panel, unlike in  
Fig.~\ref{fig:imp_mach1.4} where the scaling is kept constant.} 
\label{fig:imp_alp-1.5}
\end{figure}

Upon starting the simulation, a radiative cooling layer is impulsively
generated as the supersonic flow plows into the dense cool layer. This
set up is similar to that adopted by Sutherland \etal (\cite{SBD2003})
but with the difference that in our work the outflow boundary is much
further downstream. In Fig.~\ref{fig:imp_mach1.4} we show the
time-dependent behaviour of simulations with $M = 1.4$ and $\alpha =
-2.5$, $-2.0$, and $-1.5$.  It is clear from Fig.~\ref{fig:imp_mach1.4}
that the overstability becomes damped as $\alpha$ increases, but
exists if $\alpha$ is small enough. When
$M=1.4$, we find damping is achieved with $\alpha \gtsimm -1.8$. Comparison
with Fig.~\ref{fig:cool_alpha2} reveals that the value of $\alpha$
required for the overstability of low Mach number flows occurs within
certain limited temperature ranges.  In Fig.~\ref{fig:imp_alp-1.5} we
show the time-dependent behaviour of simulations with $\alpha=-1.5$
and $M=1.4$, 2, 3, and 5. It is clear that radiative shocks become
increasingly susceptible to overstability as $M$ increases, a finding
which is in agreement with earlier work. In Table~\ref{tab:fund_damp}
we list the critical value of $\alpha$ for damping of the fundamental mode.
Note that nonlinear effects may contribute to these values (see, \eg,
Strickland \& Blondin \cite{SB1995}). We discuss this
further in connection with a linear stability analysis of low
Mach number shocks (Pittard \etal in preparation). 

Table~\ref{tab:fund_damp} also reveals that a $M=5$ radiative
shock has a value of $\alpha_{\rm cr}$ which is still far from that
which occurs in the strong shock limit ($\alpha_{\rm cr} \approx 0.4$,
\cf Chevalier \& Imamura \cite{CI1982}). Calculations reveal that this
limit is not reached until Mach numbers of $\approx 100$, with shocks
of $M=10$, 20, and 40 having $\alpha_{\rm cr} \approx -0.1$, 0.2 and
0.3, respectively.

\begin{table}
\begin{center}
\caption{Critical value of $\alpha$ for damping of the fundamental mode, and
angular frequency (when $\alpha=-1.5$) as a function of the Mach number. 
For $\alpha > \alpha_{\rm cr}$ the 
fundamental mode is damped, though the shock may still be unstable to
higher order modes, such as the first overtone (\cf Chevalier \& Imamura
\cite{CI1982}; Strickland \& Blondin \cite{SB1995}).}
\label{tab:fund_damp}
\begin{tabular}{llc}
\hline
$M$ & $\alpha_{\rm cr}$ & $\omega_{\rm f}\;(\alpha=-1.5)$ \\   
\hline
 1.40 & -1.8 & 0.73 \\
 2.00 & -1.2 & 0.51 \\
 3.00 & -0.7 & 0.32 \\
 5.00 & -0.4 & 0.22 \\
\hline
\end{tabular}
\end{center}
\end{table}

\begin{figure}[ht]
\begin{center}
\psfig{figure=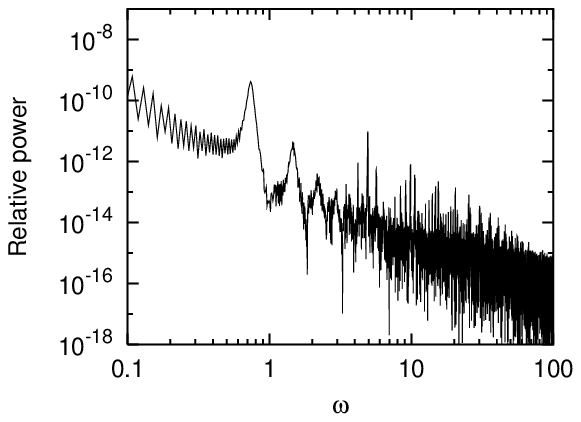,width=8.0cm}
\psfig{figure=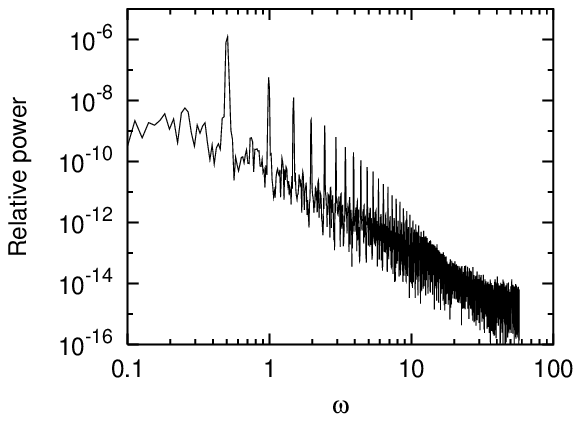,width=8.0cm}
\psfig{figure=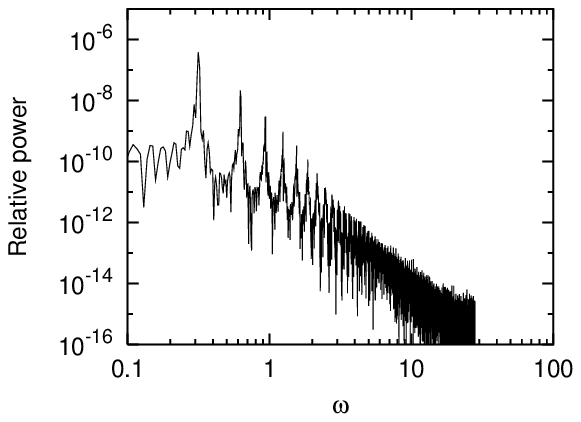,width=8.0cm}
\psfig{figure=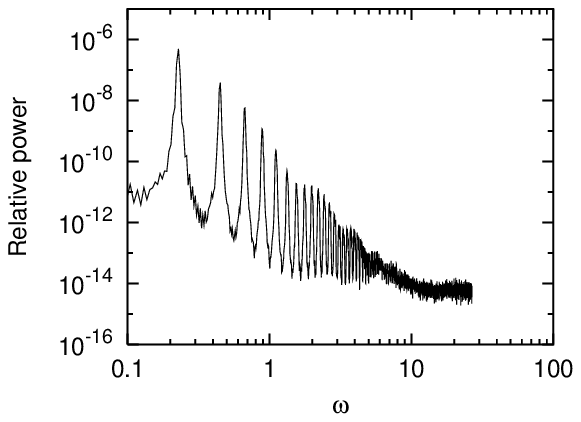,width=8.0cm}
\end{center}
\caption[]{Power spectra (mean square amplitude) for the simulations shown in 
Fig.~\ref{fig:imp_alp-1.5} where $\alpha=-1.5$. 
$M$ varies from 1.4, 2, 3, 5 from the top to bottom panels 
respectively.}
\label{fig:powerspct_alp-1.5_imp}
\end{figure}

Power spectra of the shock position of the simulations shown in
Fig.~\ref{fig:imp_mach1.4} reveal that the dimensionless angular
frequency of the fundamental mode, $\omega_{\rm f}$, is relatively
insensitive to $\alpha$, being 0.665, 0.690, and 0.734 for $\alpha$ of
$-2.5$, $-2.0$ and $-1.5$ respectively\footnote{For ease of comparison
with previous work, $\omega_{\rm f}$ is defined in units of $v_{\rm
0}/x_{\rm b}$, where $v_{\rm 0}$ is dimensionless and equal to unity
and where $x_{\rm b}$ is also dimensionless.}.  This trend is
consistent with earlier simulations of higher Mach number shocks
(Strickland \& Blondin \cite{SB1995}). However, larger changes in
$\omega_{\rm f}$ occur when $\alpha$ is kept constant and $M$ is
varied.  This is illustrated in Fig.~\ref{fig:powerspct_alp-1.5_imp}
where power spectra of the simulations shown in
Fig.~\ref{fig:imp_alp-1.5} are displayed. Table~\ref{tab:fund_damp}
also notes the value of $\omega_{\rm f}$ as a function of $M$ for
shocks with $\alpha=-1.5$. As $M$ increases, $\omega_{\rm f}$
decreases.

\subsection{Alternative set ups}
\label{sec:altern}
One alternative to the work in the previous section is to impose, at
the beginning of the simulation, a steady-state shock structure in the
middle of the grid.  The overstability may then be excited from weak
perturbations produced by the numerical mapping of the steady solution
onto the computational grid. While the initial conditions are clearly
different, the boundary conditions are identical to the previous case
(that is inflowing pre-shock gas at the upstream boundary, and
outflowing, cold, compressed gas at the downstream
boundary). Calculations based on this method are presented in
Strickland \& Blondin (\cite{SB1995}) and Pittard \etal
(\cite{P2003}).

Another possible set up is that of a reflecting wall. In this scenario,
the grid initially contains a supersonic flow, the upstream boundary is
set to free-inflow, and the downstream boundary is set to a pure reflector
(in practice this is usually achieved with ghost cells which are exact copies 
of the cells close to the boundary but with their velocity vector reversed).
As the simulation evolves a shock is launched from the reflecting boundary
and moves upstream into the oncoming flow. Shocked gas which subsequently 
cools piles up against the boundary (some earlier simulations in the 
literature do not use a ``pure'' reflecting boundary, \eg, Imamura \etal
\cite{IWD1984}).

Since the compression ratio is relatively small for low Mach number 
shocks (see Table~\ref{tab:properties}), simulations with reflecting 
conditions at the downstream boundary required a much larger
grid than the simulations with alternative set ups. This is because
in the latter cases the cold gas flows off the grid and the calculation
is performed in a reference frame where the average position of the
shock is stationary. In contrast, simulations where the 
downstream boundary is reflecting are performed in a reference frame
where this boundary is stationary. The width of the CDL continuously grows 
during such calculations as cold gas is not allowed to flow off the grid.
The growth of the CDL pushes the shock upstream into the oncoming flow. 
The normalized speed at which the CDL grows is $1/\gamma M^{2}$, and 
this must be accounted for when 
calculating the velocity of the pre-shock gas in order that a shock
with the desired Mach number is obtained.
  
In Fig.~\ref{fig:m3.0_alp-1.5_compv1} we compare time-space diagrams of
simulations with $M=3$ and $\alpha=-1.5$ and the 3 different
set ups noted above. In the top panel the overstability is self-excited
by the small numerical errors which exist from the initialization. Linear 
growth is observed for about the first 50\% of the run, after which the 
amplitude of the oscillations saturate and the system enters the nonlinear
regime. The elapsed time before the system saturates depends on the 
numerical resolution - finer grids have smaller initialization errors,
so the shock is subject to weaker perturbations at startup, and the 
timescale until saturation is longer.

The middle panel of Fig.~\ref{fig:m3.0_alp-1.5_compv1} repeats the
impulsively generated simulation shown in the third panel of
Fig.~\ref{fig:imp_alp-1.5}. In the bottom panel of
Fig.~\ref{fig:m3.0_alp-1.5_compv1} we show the evolution of the
overstability for flow running into a reflecting downstream
boundary. An overstable shock is immediately generated, and the
thickness of the CDL increases rapidly with time due to the low
compression. Unlike in the previous setups, shock waves in the
CDL are reflected off the downstream boundary, and have significant
influence on the position of the shock early in the
simulation. However, the oscillations become very regular as time
progresses, although only the first 13 oscillations are shown here. In
the simulations in the upper and middle panels of
Fig.~\ref{fig:m3.0_alp-1.5_compv1} the large distance to the
downstream boundary means that there is no mechanism for returning
waves transmitted into the CDL back into the cooling zone of the
shock.

For ease of comparison, Fig.~\ref{fig:m3.0_alp-1.5_comp} shows the
time evolution of the shock position for the simulations shown in
Fig.~\ref{fig:m3.0_alp-1.5_compv1} but where we have accounted
for the build-up of the CDL in the reflecting simulation
by subtracting the time-averaged 
growth rate in its width from the position of the shock.
Broad agreement is seen in the amplitude and frequency
of the oscillations once the simulations begin to settle down,
particularly between the upper and middle plots. The time required for
the amplitude of the oscillations to saturate in the top plot is a
function of the grid resolution, since start-up errors are reduced as
the resolution increases. As already noted, the simulation shown in the 
lower plot becomes very regular as time
progresses. A power spectrum analysis reveals that the fundamental
frequency is the same in all 3 simulations to within 1\%.

\subsection{The effect of the CDL}
\label{sec:cdl}
The simulations in the previous section lead us to conclude that for
low Mach number shocks the particular form of the boundary conditions
makes little difference to the stability. It is interesting to note
that the continued growth of the CDL does not affect the damping of
the oscillations.

\begin{figure}[ht]
\begin{center}
\psfig{figure=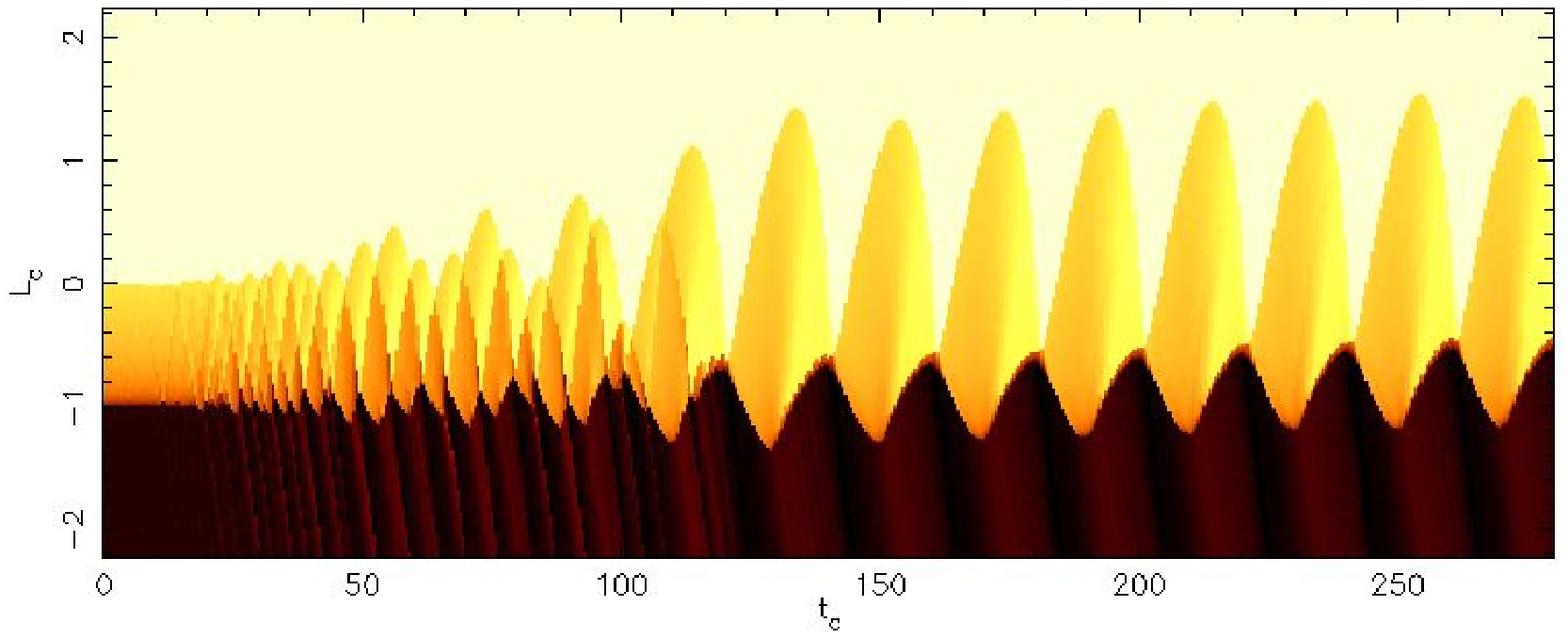,width=9.0cm,angle=0}
\psfig{figure=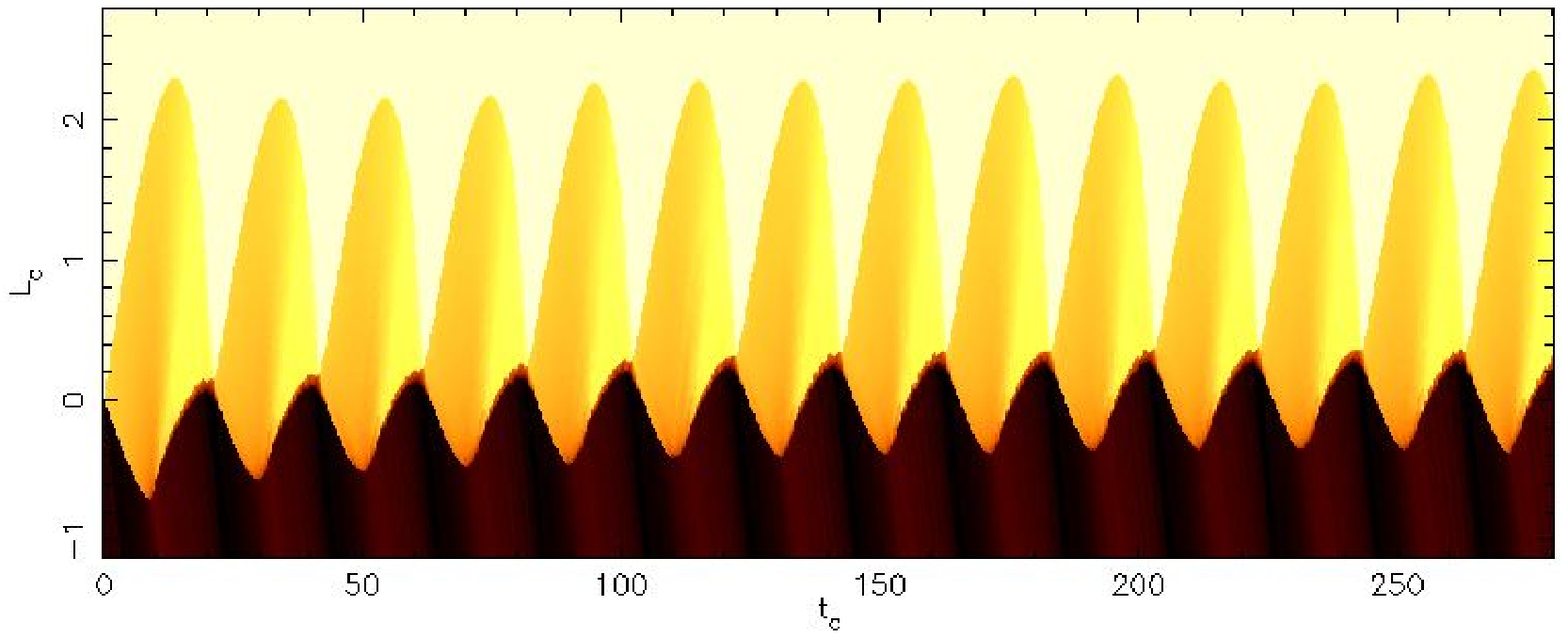,width=9.0cm,angle=0}
\psfig{figure=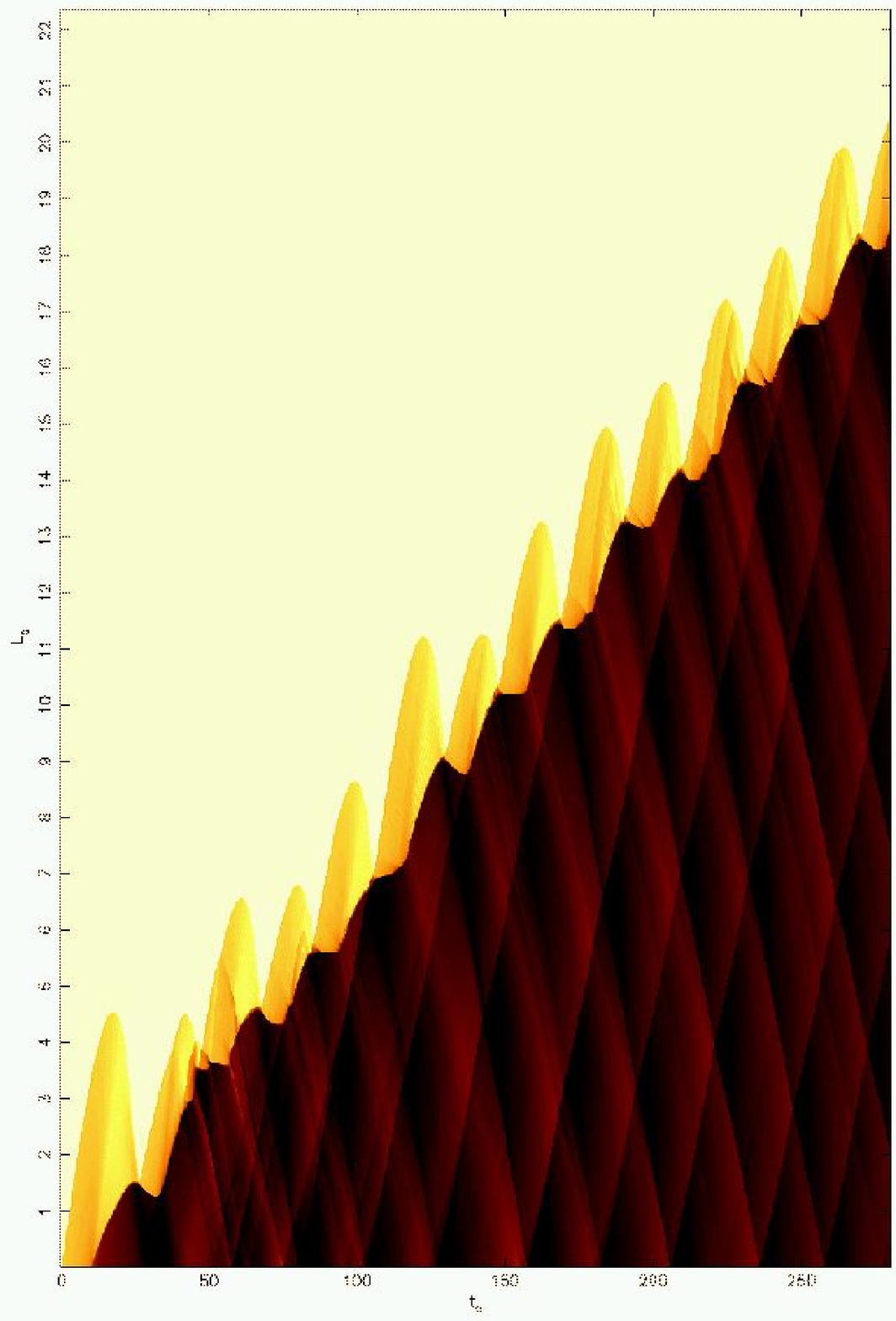,width=9.0cm}
\vspace{-15mm}
\end{center}
\caption[]{Time-space diagrams of the density evolution of a
1-dimensional radiative shock with $M=3$ and cooling exponent $\alpha=-1.5$.
In the top panel the flow is initialized to the steady state
solution. In the middle panel the shock is impulsively generated. In
the bottom panel the downstream boundary condition is reflecting.}
\label{fig:m3.0_alp-1.5_compv1}
\end{figure}

The dynamics of the CDL was extensively investigated by Walder \&
Folini (\cite{WF1996}), who concluded that the CDL can strongly
influence the stability of the cooling zone (see also Strickland \&
Blondin \cite{SB1995}). The oscillation of the CDL introduces
time-dependent conditions at the rear boundary of the cooling layer,
which may then affect the evolution of the radiative shock.
Sound waves and shocks within the CDL (see Fig.~\ref{fig:m3.0_alp-1.5_compv1})
are created by the pressure on the upstream side of the CDL varying
as the cooling region oscillates through its cycle of overstability.
The boundary between the CDL and the cooling region oscillates in
anti-phase with the shock because as the cooling region begins to lose
pressure support, the CDL finds itself overpressurized compared to
the upstream flow (and vice-versa). These anti-phase oscillations
tend to damp the overstability of the cooling region, since
movement of the CDL into the cooling region helps to maintain the pressure
in the cooling region, and vice-versa.

In those cases where the cooling layer disappears at the end of each
oscillation cycle, the behaviour of the system should approximately be
given by the solution of a Riemann problem, with the ``left'' and ``right''
states the CDL and the supersonic upstream flow, respectively. In 
Table~\ref{tab:riemann} we list the flow quantities for the ``left'' and
``right'' states of the impulsive shock generation simulations shown in
Fig.~\ref{fig:mvar_alp-1.5}, and the Riemann solution for the wave speeds
moving into these states. In each of the Riemann solutions noted in
Table~\ref{tab:riemann}, the left and right waves are both shocks, and their
velocity and that of the contact discontinuity are in good agreement with
those apparent in the numerical simulations in Fig.~\ref{fig:mvar_alp-1.5}.
The velocity of the resolved state in the Riemann solution, $v_{\rm cd}$, 
which corresponds to the velocity of the boundary between the CDL and the
cooling layer, is negative for both the $M=5$ and $M=10$ flows, and means 
that this boundary initially moves downstream as the cooling layer grows,
as shown in Fig.~\ref{fig:mvar_alp-1.5}. We also see that $v_{\rm cd}$
becomes less negative with increasing $M$, again in agreement with
Fig.~\ref{fig:mvar_alp-1.5}. Finally, the velocity of both the left and
right waves, $v_{\rm l}$ and $v_{\rm r}$, increase as $M$ increases, the
former being consistent with the shock oscillation amplitude increasing
with $M$.

\begin{table*}
\begin{center}
\caption{The initial behaviour of the $M=5$ and $M=10$ simulations
shown in Fig.~\ref{fig:mvar_alp-1.5} can be determined from the
solution to a Riemann problem, which yields the resulting wave
velocities in the system ($v_{\rm l}$, $v_{\rm cd}$, and $v_{\rm r}$ being the
velocity into the left state, the velocity of the contact discontinuity, and
the velocity into the right state).  
The left state is defined as the CDL, while
the right state is defined as the pre-shock flow (see
Fig.~\ref{fig:isoimp}). While the values quoted are for the initial
conditions, they should be a good representation of the conditions
which exist at the exact moment that the cooling layer disappears when
$\alpha << \alpha_{\rm cr}$.}
\label{tab:riemann}
\begin{tabular}{cccccccccc}
\hline
    & \multicolumn{3}{c}{Left state} & \multicolumn{3}{c}{Right state} & \multicolumn{3}{c}{Wave velocities} \\
$M$ & $\rho$ & $p$ & $v$ & $\rho$ & $p$ & $v$ & $v_{\rm l}$ & $v_{\rm cd}$ & $v_{\rm r}$ \\
\hline  
5 & 41.67 & 1.0 & -0.024 & 1.0 & 0.024 & -1.0 & -0.243 & -0.052 & 0.295 \\
10 & 166.67 & 1.0 & -0.006 & 1.0 & 0.006 & -1.0 & -0.117 & -0.022 & 0.312 \\
\hline
\end{tabular}
\end{center}
\end{table*}

\begin{figure}[ht]
\begin{center}
\psfig{figure=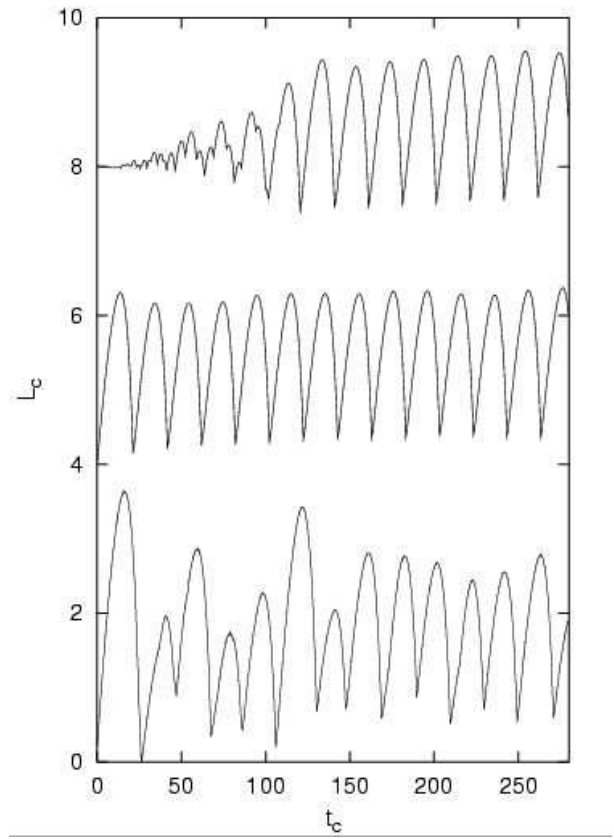,width=8.0cm}
\end{center}
\caption[]{The shock position as a function of time for simulations
with $M=3$ and $\alpha=-1.5$. The set up with an isolated steady-state
shock as the initial condition is shown at the top, the set up with
impulsive shock generation is shown in the middle, and the set up with
a reflecting downstream boundary (\ie ``flow into a wall'') is shown
at the bottom. In the latter case the growth in width of the cold
dense layer has been removed to aid comparison with the other
scenarios, and each dataset is offset with respect to the others.}
\label{fig:m3.0_alp-1.5_comp}
\end{figure}

\begin{figure}[ht]
\begin{center}
\psfig{figure=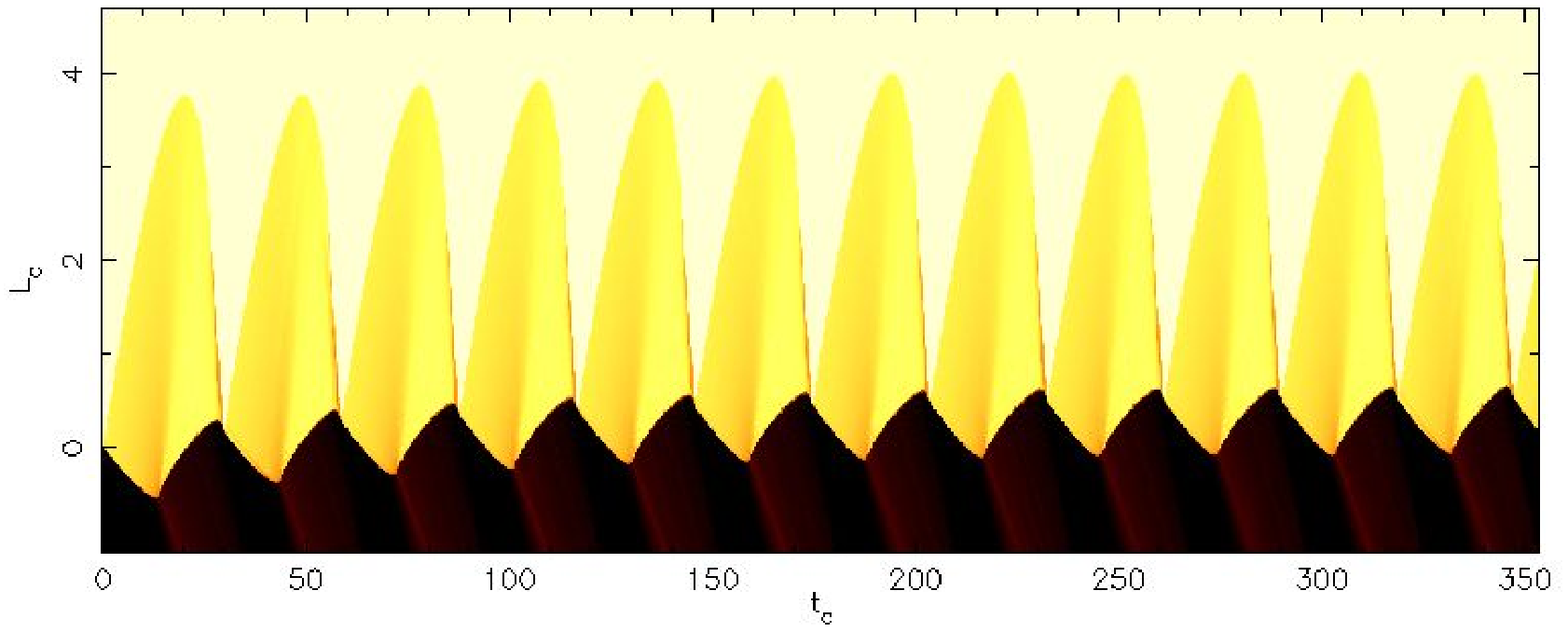,width=9.0cm,angle=0}
\psfig{figure=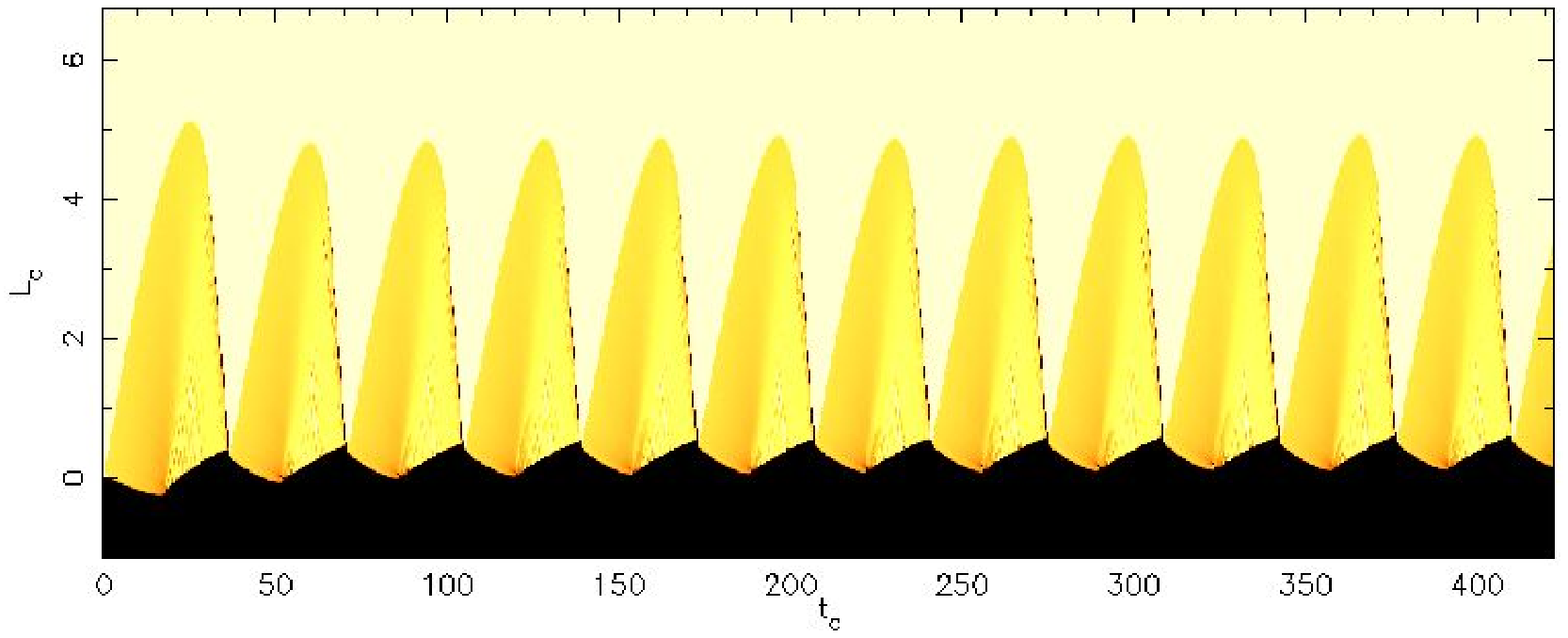,width=9.0cm,angle=0}
\end{center}
\caption[]{Time-space diagrams of the density evolution of a
1-dimensional radiative shock with cooling exponent $\alpha=-1.5$
and $M=5$ (top) and $M=10$ (bottom).}
\label{fig:mvar_alp-1.5}
\end{figure}

While the initial behaviour of the boundary between the CDL and
the cooling region can be understood in terms of the Riemann problem,
its later evolution is dependent on the behaviour of the cooling region,
and vice-versa.
In Fig.~\ref{fig:mvar_alp-1.5} we see that as the Mach number
increases the oscillation amplitude of the CDL boundary decreases
and the amplitude of the overstability increases (when specified in
units of $L_{\rm c}$). Increased damping of the overstability occurs
when the oscillation amplitude of the CDL boundary increases, and such
feedback between the CDL and the cooling gas is clearly dependent on
the Mach number of the shock. In the general case (\ie when the
cooling region does not necessarily completely disappear between
oscillations), the degree of feedback between the CDL and the cooling
gas can be quantified by considering how much wave
energy from the oscillating post-shock flow is transmitted into the
CDL, and how much is reflected. We assume that to first order we can
represent the boundary between the cooling post-shock gas and the CDL
as a discontinuous boundary between two states with densities
$\rho_{\rm s}$ (the post-shock density) and $\rho_{\rm c}$ (the
density of the CDL). The reflection coefficient is then (Landau \&
Lifshitz \cite{LL1984})
\begin{equation}
\label{eq:reflect}
R = \left(\frac{\rho_{\rm c}c_{\rm c} - \rho_{\rm s}c_{\rm s}}{\rho_{\rm c}c_{\rm c} + \rho_{\rm s}c_{\rm s}}\right)^{2},
\end{equation}
where $c_{\rm c}$ and $c_{\rm s}$ are the sound speeds in the CDL and
immediately post-shock, respectively. Evaluating Eq.~\ref{eq:reflect}
reveals that $R = 0.074$, 0.11, 0.20, 0.36, and 0.59 for $M=1.4$, 2,
3, 5, and 10. Thus, as the Mach number increases, an increasing
percentage of the incident wave energy is reflected back into the cooling
region. This is consistent with a decreased oscillation amplitude with 
increasing Mach number of the boundary between the CDL and cooling zone 
(see Fig.~\ref{fig:mvar_alp-1.5}), since less energy is
transmitted into it. It is clear, therefore, that the CDL is more able
to damp the shock overstability when the Mach number is small, since
it is then more efficient at absorbing the energy of the
oscillations. This efficiency is independent of the thickness 
of the CDL, and explains why we do not see more efficient damping, indicated
by a decrease in the oscillation amplitude of the cooling region, when
the thickness of the CDL grows (see Figs.~\ref{fig:m3.0_alp-1.5_compv1} 
and~\ref{fig:m3.0_alp-1.5_comp}). At high Mach numbers the 
fraction of transmitted wave energy approaches zero, the velocity of the
CDL boundary and the amplitude of its
oscillations become very small, and feedback
between the CDL and the cooling region vanishes. The Mach number
dependent transmission efficiency of sound wave energy is undoubtedly 
the reason why we observe a strong Mach number dependence of
$\alpha_{\rm cr}$ and $\omega_{\rm f}$ at low Mach numbers, but not at
high Mach numbers (where $R$ reaches an asymptotic value of 1.0).

Our results are consistent with those presented by
Walder \& Folini (\cite{WF1996}), who noted from their studies of
shocks with higher Mach numbers that feedback between the CDL and the
cooling region is relatively minor if the pre-shock gas is smooth (in
agreement with our $M=10$ simulations). They also found that the feedback is
stronger if the
shock encounters a disturbance in the pre-shock medium. In such
situations the CDL may oscillate with significantly greater amplitude,
which in turn may cause the overstability to increase or decrease in
amplitude and/or frequency, and even switch on and switch off (see
Figs.~15 and~16 in Walder \& Folini \cite{WF1996}).
While we have not explored the effect of such upstream disturbances
in this work, we anticipate that these changes in behaviour will
apply equally to low Mach number shocks.

An interesting point in our simulations is that energy transmitted
into the CDL can leave it in two different ways.  Waves in the
CDL will be damped through radiative losses as compressions increase
the temperature above $T_{\rm 0}$. In addition, energy may be lost
from the system if the downstream boundary on the hydrodynamical grid
has outflow conditions (though no such losses occur in the simulations
where we impose a reflecting wall at the back of the CDL). In the
simulations noted in Sec.~\ref{sec:iso_imp} the downstream boundary is
far enough away that the transmitted waves never reach it.

We also note that as the CDL is thick behind low Mach number
shocks, such shocks should be less susceptible to other
types of instability than their higher Mach number counterparts, 
such as the non-linear thin shell instability
(Vishniac \cite{V1994}; Blondin \& Marks \cite{BM1996}), and
non-radial instabilities.
 
\subsection{Cooling below the pre-shock temperature}
\label{sec:tcdl_ne_tamb}
In all of the simulations in the previous sections, the shocked gas 
is prevented from cooling below the temperature of the pre-shock gas.
Since there is no fundamental reason why this could not occur,
we have examined the nature of the 
overstability when $\chi = T_{\rm c}/T_{0} < 1$.

\begin{figure}[ht]
\begin{center}
\psfig{figure=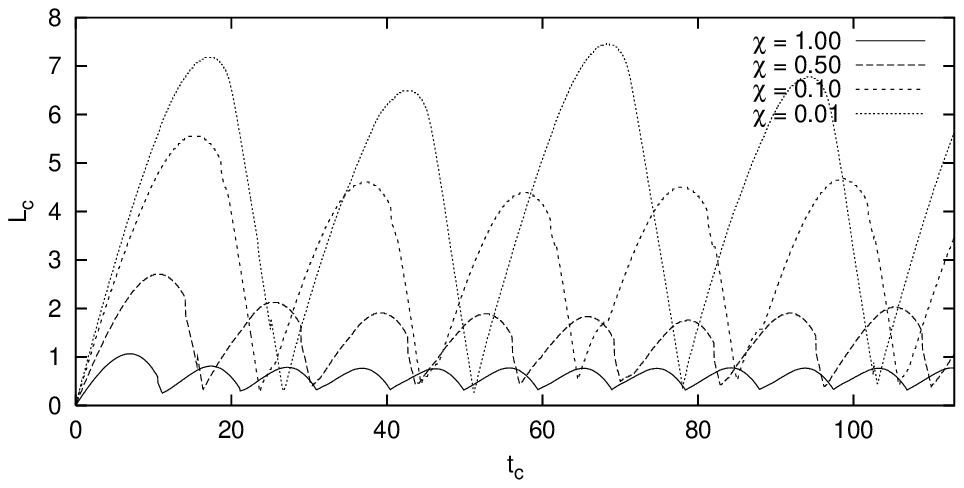,width=8.0cm,angle=0}
\end{center}
\caption[]{The time-evolution of the shock position as a function of $\chi$ 
when $M=1.4$ and $\alpha=-2.5$.}
\label{fig:beta_var1}
\end{figure}

\begin{figure}[ht]
\begin{center}
\psfig{figure=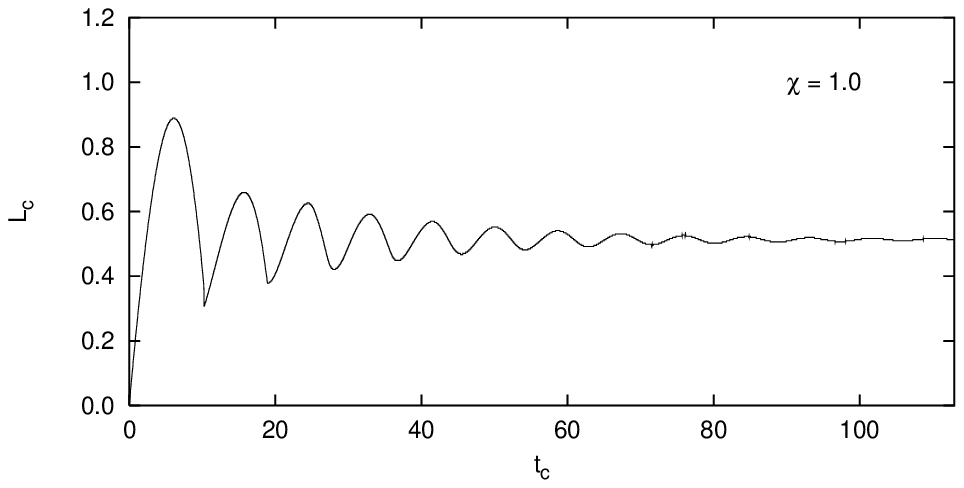,width=8.0cm,angle=0}
\psfig{figure=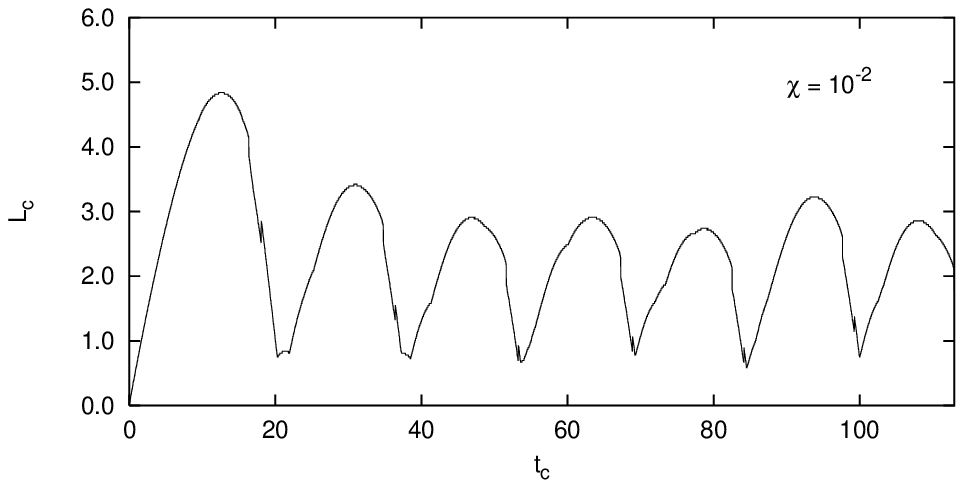,width=8.0cm,angle=0}
\end{center}
\caption[]{The shock position as a function of time when $M=1.4$ and
$\alpha=-1.5$. The ratio of the temperature of the CDL to the
pre-shock temperature, $\chi$, is noted in each panel.}
\label{fig:beta_var2}
\end{figure}

We find that the value of $\chi$ has a dramatic effect on the
properties of low Mach number shocks. Fig.~\ref{fig:beta_var1} shows
the time evolution of the position of the shock front as a function of
$\chi$ when $M=1.4$ and $\alpha=-2.5$. It is obvious that decreasing
$\chi$ causes the oscillation amplitude to increase and the frequency
of the fundamental mode to decrease, which is the same trend obtained
when $M$ increases with $\chi=1$ and $\alpha$ fixed. Clearly the value
of $\chi$ has a strong influence on the behaviour of the shock, though
we see convergence at low values of $\chi$ (the behaviour of a
simulation with $M=1.4$, $\alpha=-2.5$, and $\chi=10^{-3}$, is not too
different from that with $\chi=10^{-2}$). At high Mach numbers (\eg,
$M=40$), varying $\chi$ has a negligible effect.
Fig.~\ref{fig:beta_var2} shows that the critical value of $\alpha$ for
stability is also dependent on $\chi$. In the top panel the
overstability is damped, but this is not the case in the lower
panel. In Table~\ref{tab:alpha_cr_chi} we note values of $\alpha_{\rm
cr}$ as a function of both $M$ and $\chi$. There are large changes in
$\alpha_{\rm cr}$ with $\chi$ at low Mach numbers, with the result
that it is much easier for a low Mach number shock to be overstable
when $\chi < 1$. We do not obtain $\alpha_{\rm cr}=0.4$ at lower 
Mach numbers when $\chi$ is reduced.

\begin{table}
\begin{center}
\caption{Critical value of $\alpha$ for damping of the fundamental mode 
as a function of the Mach number and $\chi$, the ratio of the temperature
of the CDL to the pre-shock temperature. This table is an extension of
Table~\ref{tab:fund_damp} where $\alpha_{\rm cr}$ is listed for $\chi=1$.}
\label{tab:alpha_cr_chi}
\begin{tabular}{l|llll}
\hline
$M$ & \multicolumn{4}{c}{$\chi$} \\
    & 1.0 & 0.5 & 0.1 & 0.01 \\   
\hline
 1.40 & -1.8 & -0.6 & -0.2 & -0.1 \\
 2.00 & -1.2 & -0.6 & -0.2 & -0.1 \\
 3.00 & -0.7 & -0.5 & -0.2 & -0.1 \\
 5.00 & -0.4 & -0.3 & -0.1 & -0.1 \\
10.0  & -0.1 & -0.1 &  0.1 &  0.3 \\
20.0  &  0.2 &  0.2 &  0.2 &  0.3 \\
40.0  &  0.3 &  0.3 &  0.3 &  0.3 \\
100   &  0.4 &  0.4 &  0.4 &  0.4 \\ 
\hline
\end{tabular}
\end{center}
\end{table}

It is clear from Figs.~\ref{fig:beta_var1} and~\ref{fig:beta_var2} and
Table~\ref{tab:alpha_cr_chi} that reducing $\chi$ below unity makes
the shock behave like a higher Mach number shock with $\chi=1$. This
is expected since if $\chi < 1$, the density of the CDL
increases above the value given by $\gamma M^{2}$ (the cooling
time and length of the shocked gas also increase, though by amounts
which are dependent on $\alpha$). A shock with
$M=1.4$ and $\chi=0.1$ has $\rho_{\rm c}/\rho_{0} = 42$, which is
almost identical to a shock with $M=5$ and $\chi=1$. However, the
value of the reflection coefficient, $R = 0.57$, is almost equivalent
to that from a shock with $M=10$ and $\chi=1$. It is not surprising,
therefore, that the value of $\alpha_{\rm cr}$ for a shock with
$M=1.4$ and $\chi=0.1$ is in good agreement with the value obtained
for a shock with $M=10$ and $\chi=1$.  Thus low Mach number shocks
with $\chi < 1$ are more susceptible to overstability than a
straightforward comparison of $\rho_{\rm c}/\rho_{0}$ would
suggest. An evaluation of $R$ in this case allows a much more accurate
estimate of $\alpha_{\rm cr}$.

\begin{figure}[ht]
\begin{center}
\psfig{figure=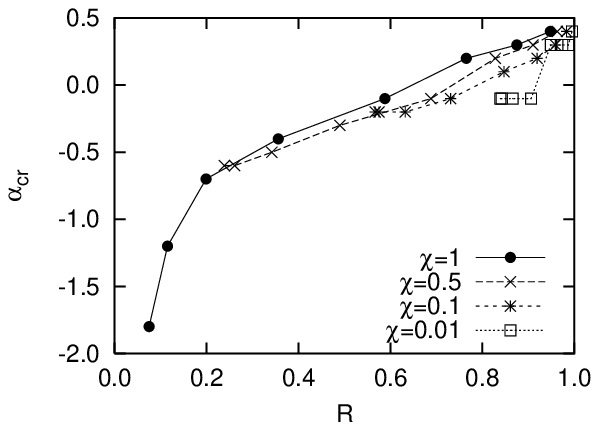,width=8.0cm,angle=0}
\end{center}
\caption[]{The value of $\alpha_{\rm cr}$, the critical value for
damping of the overstability, as a function of the reflection 
coefficient, $R$, for four different values of $\chi$. The
lines are only meant to guide the eye.}
\label{fig:r_vs_alphacr}
\end{figure}

Values of $R$ as a function of $M$ and $\chi$ are shown in
Table~\ref{tab:r_coeff}. In Fig.~\ref{fig:r_vs_alphacr} we display the
value of $\alpha_{\rm cr}$ as a function of $R$ for 4 different values
of $\chi$. It is clear from the relatively tight relation between the
curves that the value of $R$ can be used to obtain a reasonable
estimate of $\alpha_{\rm cr}$. The larger scatter in $\alpha_{\rm cr}$
for a given $R$ which exists in the top right of the plot indicates
that our simple method of evaluating $R$ is a somewhat poorer
description of the actual dynamics in this region of parameter space.

We also find that not all of the shock properties scale in such a
straightforward manner. For example, the oscillation frequency of the
fundamental mode when $M=1.4$, $\alpha=-2.5$, and $\chi=0.1$ is
$\omega_{\rm f} = 0.31$, which is some way from the value $\omega_{\rm
f} = 0.14$ obtained when $M=10$, $\alpha=-2.5$, and $\chi=1$ (in
comparison, when $M=1.4$, $\alpha=-2.5$, and $\chi=1$, we find
$\omega_{\rm f} = 0.67$). Thus the actual Mach number of the shock is
still important.

\begin{table}
\begin{center}
\caption{Reflection coefficient, $R$, as a function of the Mach number 
and $\chi$, the ratio of the temperature
of the CDL to the pre-shock temperature.}
\label{tab:r_coeff}
\begin{tabular}{l|llll}
\hline
$M$ & \multicolumn{4}{c}{$\chi$} \\
    & 1.0 & 0.5 & 0.1 & 0.01 \\   
\hline
 1.40 & 0.075 & 0.239 & 0.567 & 0.839 \\
 2.00 & 0.115 & 0.261 & 0.575 & 0.842 \\
 3.00 & 0.199 & 0.342 & 0.632 & 0.866 \\
 5.00 & 0.356 & 0.490 & 0.731 & 0.906 \\
10.0  & 0.588 & 0.688 & 0.847 & 0.949 \\
20.0  & 0.765 & 0.828 & 0.919 & 0.974 \\
40.0  & 0.875 & 0.910 & 0.959 & 0.987 \\
100   & 0.948 & 0.963 & 0.983 & 0.995 \\ 
\hline
\end{tabular}
\end{center}
\end{table}

\section{Discussion}
\label{sec:discuss}
Radiative shocks occur ubiquitously in astronomy, \eg, in
SNRs, stellar wind bubbles, and jets
from young-stellar-objects. They exist over a wide range of
spatial scales, from accreting white dwarfs 
to galactic scale-accretion.
Evidence consistent with the presence of an unsteady flow, such as
produced by the radiative overstability, has been found in
observations of the Vela SNR (Raymond \etal \cite{RWB1991}) - 
unusually broad line widths (of Si{\sc ii} and Mg{\sc ii}) and differences 
between the shock ram pressure and the postshock thermal pressure.
The use of steady shock models in such circumstances means that 
conclusions drawn from the observed line ratios will be incorrect
(Innes \etal \cite{IGF1987a},~\cite{IGF1987b}; Gaetz \etal \cite{GEC1988}).

A key point concerning SNRs is that the deceleration of the blast wave
limits the number of oscillations that may occur, since there is only
a finite time during which the shell is unstable (see, \eg, Mac Low \&
Norman \cite{MN1993}; Blondin \etal \cite{BWBR1998}). As noted in
Sec.~\ref{sec:intro}, previous works have assumed that the temperature
of the ISM corresponds to the WIM phase (\ie $T_{\rm ISM} \approx
10^{4}\;$K; see McKee \& Ostriker \cite{MO1977}). The consequence of this
is that low Mach number shocks have post-shock temperatures which are
on the steeply rising part of the cooling function, and will thus
be stable. However,
the ISM is known to consist of a variety of phases, with
its volume likely dominated by gas with a temperature of $\sim
10^{5}-10^{6}$K (the HIM phase in the nomenclature of 
McKee \& Ostriker \cite{MO1977}). This hot
phase is continuously regenerated through the combined actions of
supernova explosions and the winds of massive stars. SNRs 
interacting with this phase of the ISM have much reduced Mach numbers.

The evolution of SNRs can be broken up into a sequence of stages.
From an initial ejecta-dominated stage, they may evolve through the
Sedov-Taylor (ST), pressure-driven snowplough (PDS), and
momentum-driven snowplough (MDS) stages. Once the SNR is in the PDS
stage, its forward shock may be susceptible to the radiative
overstability. The timescale for the transition between the ST and PDS
stages is roughly (Blondin \etal \cite{BWBR1998}):
\begin{equation}
\label{eq:t_rad}
t_{\rm rad} \approx 2.9 \times 10^{4} E_{51}^{4/17}n_{0}^{-9/17} \;{\rm yr},
\end{equation}
where $E_{51}$ is the kinetic energy of the original explosion in
units of $10^{51}\;$ergs, and $n_{0}$ is the number density of the
ISM.  The velocity of a ST blast wave at this age is $v_{\rm
s} \approx 260 E_{51}^{1/17}n_{0}^{2/17}\;\kmps$. The sound speed in 
the hot ISM is $c \approx 150 T_{\rm ISM,6}^{1/2}\;\kmps$, where 
$T_{\rm ISM,6}$ is the ISM temperature in units of $10^{6}\;$K. 

For the blast wave to be susceptible to overstability when the SNR is
expanding at relatively low Mach numbers requires that the post-shock
gas is located on a part of the cooling curve where $\alpha <
\alpha_{\rm cr}$. The simulations in Sec.~\ref{sec:hydro} show that
$\alpha \ltsimm -0.7$, $-1.2$, and $-1.8$ is needed in order to have
overstable shocks at Mach numbers of 3, 2, and 1.4, respectively.
Such values of $\alpha_{\rm cr}$ are fairly demanding in the sense
that $\alpha$ is less than $\alpha_{\rm cr}$ only in very limited
temperature ranges. Nonetheless, SNRs evolving into an ISM with
$T_{\rm ISM} \sim 10^{5}\;$K or $T_{\rm ISM} \gtsimm 10^{6}\;$K may
satisfy these conditions (see Fig.~\ref{fig:cool_alpha2}). However,
the work in Sec.~\ref{sec:tcdl_ne_tamb} showed that the criteria for
overstability becomes far less stringent if the post-shock gas cools
to a lower temperature than the pre-shock gas ($\alpha_{\rm cr}
\approx -0.1$ when $\chi=0.1$ for $M \geq 1.4$), in which case SNR
shocks are readily overstable.  Table~\ref{tab:snrproperties} shows
values for $t_{\rm rad}$, $v_{\rm s}$, $T_{\rm s}$ and $M$ for two
values of $T_{\rm ISM}$ and for different values of $n_{0}$, assuming
$E_{51}=1$ and $\gamma=5/3$.  Since $t_{\rm tr}$ is an approximation
to the time of shell formation the numbers in
Table~\ref{tab:snrproperties} are only illustrative. Nevertheless,
Table~\ref{tab:snrproperties} shows that SNRs evolving in high
pressure environments may have low Mach number shocks which are
susceptible to the radiative overstability, as determined from
the slope of the cooling curve between $T_{\rm ISM}$ and $T_{\rm s}$
and by the Mach number dependence of $\alpha_{\rm cr}$ (see
Table~\ref{tab:alpha_cr_chi}). Our work in this paper is therefore
relevant to such extreme settings as starburst galaxies, where the
range of density and temperature values in
Table~\ref{tab:snrproperties} are comparable to observed ranges.

\begin{table}
\begin{center}
\caption{Approximate properties of SNR blast waves at the transition into the
pressure-driven snowplough stage for different values of $T_{\rm ISM}$
and $n_{0}$, chosen to demonstrate the
conditions under which SNRs in the pressure-driven snowplough stage
have low Mach number forward shocks which are susceptible to the
radiative overstability. The age of the SNR at the onset of shell
formation, $t_{\rm tr}$, the shock velocity,
$v_{\rm s}$, and the Mach number, $M$, at this time, are given. The values
illustrate that overstable low Mach
number radiative shocks in SNRs will exist.
For $T_{\rm ISM} = 10^{6}\;$K and $n_{0} = 3.5 \times
10^{-3}\;\pcm3$ the blast wave becomes subsonic before transition to
the pressure-driven snowplough stage.}
\label{tab:snrproperties}
\begin{tabular}{llllll}
\hline
$T_{\rm ISM}$ & $n_{0}$ & $t_{\rm tr}$
& $v_{\rm s}$ & $T_{\rm s}/T_{\rm ISM}$ & $M$ \\
(K) & $(\pcm3)$ & ($10^{5}$~yr) & $(\kmps)$ & & \\
\hline
$2 \times 10^{5}$ & $10^{-3}$ & 11.2 & 115 & 1.7 & 1.7 \\
                  & $10^{-2}$ &  3.3 & 151 & 2.4 & 2.3 \\
                  & $10^{-1}$ &  1.0 & 198 & 3.6 & 3.0 \\
                  & 1 & 0.3 & 260 & 5.6 & 3.9 \\
$10^{6}$ & $10^{-3}$ & 11.2 & 115 & - & 0.8 \\
         & $10^{-2}$ &  3.3 & 151 & 1.0 & 1.0 \\
         & $10^{-1}$ &  1.0 & 198 & 1.5 & 1.5 \\
         & 1 & 0.3 & 260 & 1.7 & 1.7 \\
\hline
\end{tabular}
\end{center}
\end{table}

It has recently become apparent that radiative shocks in dense
molecular clouds can also be unstable (Smith \& Rosen
\cite{SR2003}). While the simulations presented in this work are of
relatively high Mach number shocks, slower shocks with Mach numbers of
about 5 and post-shock temperatures of about $10^{3}\;$K, should also
be overstable (see Fig.~2a in Smith \& Rosen \cite{SR2003}), though as
usual there is the caveat that magnetic fields tend to stabilize the
overstability (T\'{o}th \& Draine \cite{TD1993}).


\section{Summary}
\label{sec:summary}
In this paper we have examined the stability of low Mach number
radiative shocks. We find that such shocks may be overstable for
sufficiently low values of $\alpha$, and that there are temperature
regions in representative cooling curves where the required values of
$\alpha$ are obtained. We have determined the critical value of
$\alpha$ at the boundary between stability and overstability as a
function of Mach number, finding that $\alpha_{\rm cr}$ increases with
$M$. The strong shock limit of $\alpha_{\rm cr} \approx 0.4$ for
damping of the fundamental mode is not reached until $M \approx 100$.
The frequency and amplitude of the fundamental mode of oscillation are
strongly dependent on the Mach number of low Mach number shocks. 
Feedback between the cooling region and the CDL is also a function of
the Mach number, and may be quantified in terms of the reflection
coefficient of sound waves in the general case. In those cases where
the cooling layer completely disappears at the end of each oscillation
cycle, the velocity of the shocks driven upstream into the pre-shock flow
and downstream into the CDL, and the velocity of the boundary between the
CDL and the cooling layer, are in good agreement with those determined
from a solution to the Riemann problem.

An important finding is that the
stability properties of low Mach number shocks are dramatically
altered if the shocked gas is able to cool below the temperature of
the pre-shock gas. In such circumstances, low Mach number shocks
behave in many ways as if they had higher Mach numbers. An increase in
$\alpha_{\rm cr}$ is the most fundamental of these changes, meaning that
such shocks are more likely to be overstable.  An investigation into
the effects of differing conditions for the initial set up and the
grid boundaries reveals that these have very little influence
on the stability criteria and oscillation frequencies. This is due to
the fact that in low Mach number shocks the CDL acts much more like a
cushion than a reflector, absorbing a substantial amount of the
incident sound wave energy. Growth in the thickness of the CDL does
not enhance the damping of the oscillations.

Our work is relevant to the evolution of SNRs interacting with the hot 
phase of the ISM (\eg, in starbursts), and to shocks in dense 
molecular clouds.

\begin{acknowledgements}
We would like to thank the referee for a constructive and helpful
report which led to further insight into this work.  JMP would like to
thank PPARC for the funding of a PDRA position and the Royal Society
for financial support.  This research has made use of NASA's
Astrophysics Data System Abstract Service.
\end{acknowledgements}

\end{document}